\documentclass[12pt,a4paper]{article}
\usepackage[utf8]{inputenc}

\usepackage{amsmath} 
\allowdisplaybreaks[4]
\usepackage{float}
\usepackage{latexsym}
\usepackage{amsfonts}       
\usepackage{amsthm} 
\usepackage{bbding}         
\usepackage{bm}             
\usepackage{graphicx}    
\usepackage[labelfont=bf]{caption}
\usepackage{multicol}
\usepackage{multirow}
\usepackage{hhline}
\usepackage{caption}
\DeclareCaptionLabelSeparator{twospace}{\ ~}   
\usepackage{subcaption}
\usepackage{fancyvrb}       
\usepackage{dcolumn}        
\usepackage{booktabs}       
\usepackage{paralist}       
\usepackage[msc-links]{amsrefs}
\usepackage{hyperref}

\usepackage{chngcntr}
\counterwithin{table}{section}
\counterwithin{figure}{section}

\usepackage{authblk}
\title{An SEIR network epidemic model with manual and digital contact tracing allowing delays}
\author[1]{Dongni Zhang}
\author[1]{Tom Britton}
\affil[1]{Department of Mathematics, Stockholm University, 106 91 Stockholm, Sweden.}
\date{\today}

\begin{document}

\maketitle

\begin{abstract} 
    We consider an SEIR epidemic model on a network also allowing random contacts, where recovered individuals could either recover naturally or be diagnosed. Upon diagnosis, manual contact tracing is triggered such that each infected network contact is reported, tested and isolated with some probability and after a random delay. Additionally, digital tracing (based on a tracing app) is triggered if the diagnosed individual is an app-user, and then all of its app-using infectees are immediately notified and isolated. The early phase of the epidemic with manual and/or digital tracing is approximated by different multi-type branching processes, and three respective reproduction numbers are derived. The effectiveness of both contact tracing mechanisms is numerically quantified through the reduction of the reproduction number. This shows that app-using fraction plays an essential role in the overall effectiveness of contact tracing. The relative effectiveness of manual tracing compared to digital tracing increases if: more of the transmission occurs on the network, when the tracing delay is shortened, and when the network degree distribution is heavy-tailed. For realistic values, the combined tracing case can reduce $R_0$ by 20-30\%, so other preventive measures are needed to reduce the reproduction number down to $1.2-1.4$ for contact tracing to make it successful in avoiding big outbreaks.
\end{abstract}

\section{Introduction}\label{sec:intro}

Contact tracing stands out as one of the most effective measures for controlling a pandemic. Its primary aim is to identify and interrupt the transmission chain by detecting who is spreading the infection. In this paper, we consider two contact tracing strategies: manual and digital contact tracing. The conventional (or so-called manual) contact tracing is usually performed by public health agencies, or else by self-reporting. Confirmed cases will be interviewed and asked to report potential infectious contacts, which will then be notified and tested. It has been recently argued that manual contact tracing alone might not have effectively controlled the COVID-19 epidemic \cite{ferretti_quantifying_2020}. Subsequently, contact tracing apps were introduced in certain countries \cites{wymant_epidemiological_2021,app_netherlands,elmokashfi2021norway,vogt2022australia}. This digital tracing approach enhances the speed of tracing by swiftly identifying and notifying the infected contacts. Modelling studies have showed that, compared to conventional tracing approaches, digital contact tracing could have a higher effectiveness due to its inherent speed \cite{kretzschmar2020impact} . When an app-user is diagnosed, immediate notifications are sent out to all of the app-users who have been recently in close contact with the confirmed case, recommending them to isolate and test themselves. Notably, contacts reached by manual tracing are mainly within the social circle, while digital tracing can also identify more anonymous contacts, e.g.\ random contacts occurring on a bus or in a store. 

 There have been studies of modelling conventional contact tracing \cites{ball2011threshold,ball2015stochastic,muller2016effect} and digital tracing \cites{ferretti_quantifying_2020,kretzschmar2020impact}. This paper aims to investigate the effectiveness of combining both manual and digital tracing and the situation when only one is in place. In addition to the random infectious contacts emphasized in most studies, our model also considers transmission through a social network. Modelling contact tracing is mathematically challenging (see, e.g., an overview in \cite{muller2021contact}). This paper focuses on the initial phase of the epidemic in a large population. The early phase of an epidemic can often be approximated by a branching process of independent individuals \cite{ball_strong_1995}. However, the introduction of contact tracing unfortunately disrupts the independence, necessitating other modelling approaches.

More specifically, we consider an SEIR epidemic spreading on a network also allowing for random contacts \cites{keeling_networks_2005,ball2008network}; that is, individuals can be infected by their network neighbours as well as by individuals randomly chosen from the whole population. We then assume that manual tracing \textit{only} happens for contacts on the network (random contacts are usually unknown), whereas digital tracing is amongst both network and random contacts, but only when both parts are app-users. Further, we introduce a \textit{tracing delay} (between diagnosis and contacts being traced) for manual tracing, of which the effect has been analysed in \cites{muller2016effect,kretzschmar2020impact}; digital contact tracing is, in contrast, triggered instantaneously. In reality, digital tracing may also encounter some tracing delays, but these are generally negligible compared to the manual tracing delays \cite{wilmink2020real}. Intuitively, longer tracing delays reduce the effectiveness of contact tracing. This is because infected individuals have more time to potentially transmit the disease before being identified and isolated.

For mathematical tractability, we assume the duration of the infectious period to be deterministic and limit both types of contact tracing to be forward and one-step only. That is, it is only possible for infectors to name infectees (and not its infector which happened earlier), and traced individuals who test positive are isolated but do not report their contacts. The simplification is inspired by \cite{ball2015stochastic}, wherein it was assumed in a uniformly mixing population, and only manual tracing is considered. Moreover, for practical purposes, we assume a configuration model for the contact network, although real-world social networks often exhibit clustering.

The structure of this paper is as follows. Section \ref{sec:model} outlines the network SEIR epidemic model and the models incorporating manual and/or digital contact tracing. In Section \ref{sec:early_stage_approx}, the early stage of the epidemic with manual and/or digital tracing is analysed via multi-type branching process approximation, and the corresponding reproduction numbers are derived. The effectiveness of combining manual and digital tracing as well as their individual effectiveness is numerically investigated in Section \ref{sec:numerical}. Conclusion and discussion are presented in Section \ref{sec:discussion}. 

\section{Model Description}\label{sec:model}
\subsection{The SEIR network epidemic model with global contacts}
\label{sec:epi_model}

We consider an \textbf{SEIR} (\textbf{S}usceptible $\rightarrow$ \textbf{E}xposed $\rightarrow$\textbf{I}nfectious$\rightarrow$\textbf{R}ecovered) epidemic model within a population of fixed size $n$ (assumed to be large), where infectious individuals establish local contacts with neighbours on a social network $G$, as well as random global contacts with individuals throughout the entire population. 

The contact network $G$ is characterised by a configuration model (\cite{danon2011networks};\cite{newman2018networks} part III, chapter 12), which features an arbitrary degree distribution $D\sim\{p_k \}$ ($\mathbf{P}(D=k)=p_k$, $k=0,1,2,...$) with finite mean $\mu$ and variance ${\sigma}^2$. Initially, one randomly selected index case is infectious, while the remainder of the population is susceptible. An infectious individual makes \textit{local} contact with each of his/her neighbours in $G$ randomly in time through independent Poisson processes with rate $\beta_{L}$. Moreover, this infective makes \textit{global} contact with other individuals (chosen randomly from the entire community) at a rate $\beta_{G}$. If the contacted individual is not susceptible, no action occurs. If susceptible, the contacted individual becomes exposed (i.e., latent) and remains latent for a random duration $T_L$, whose distribution is arbitrary but specified. A latent individual becomes infectious at the end of $T_L$. An infective remains infectious for a constant period $T_{I} \equiv \tau_{I}$, then becomes diagnosed with probability $p_D$; otherwise, the infective naturally recovers. Once diagnosed or recovered naturally, individuals play no further role in the epidemic. All contact processes and the random variables describing $T_L$ and $D$ are assumed to be mutually independent. The epidemic stops when no latent or infectious individuals remain in the population.

\subsection{Introducing manual and digital contact tracing in the model}\label{sec:comb_model}
In the following, we incorporate manual and digital contact tracing into the epidemic model defined in Section \ref{sec:epi_model}. To initiate digital contact tracing, we assume a fraction $\pi_{A}$ of individuals use a contact tracing app and follow  the recommendations to isolate themselves. 

The non-app-users can only be involved in manual contact tracing, described as follows: Once diagnosed, infected individuals are interviewed and report each of their \textit{infectee neighbours} independently with probability $p_M$. Each reported neighbour is then tested after an independent delay time having distribution $T_D$ having an arbitrary but specified distribution. If testing positive the individual immediately isolate and stop spreading the infection. 

App-users, on the other hand, are subject to both digital and manual contact tracing. If an app-user is diagnosed, all the app-using infectees are immediately notified, tested, and isolated if infectious (stopping further transmission). Meanwhile, every non-app-using infectee neighbour is reported and traced with probability $p_M$ independently with independent delays $T_D$, just like for non-app-users. For mathematical convenience, we assume that neither manual nor digital contact tracing are iterated. That is, the traced and tested individuals of a diagnosed individual will not be further contact traced. All contact processes, reporting processes, and the random variables describing $T_{D}$ and $T_{L}$ are assumed to be mutually independent. Table \ref{tab:parameter} lists all the model parameters. 

Setting $\pi_{A}=0$ gives the \textit{epidemic model with manual tracing only}, which is similar to the one introduced in \cite{ball2015stochastic}, but the epidemic there spreads in a homogeneous mixing population. If we further set $\beta_{L} = 0$, then manual tracing has no effect, which reduces to the usual SEIR homogeneous mixing epidemic model. Setting $p_{M}=0$ yields the \textit{epidemic model with digital tracing only}.

\begin{table}[htb!]
\centering
\caption{List of model parameters and random variables in the SEIR network epidemic model with manual and digital contact tracing.}
\begin{tabular}{|l|l|}
\hline
\textbf{Variable} & \textbf{Description} \\
\hline
$D \sim\{p_k \}$ & degree distribution of configuration-type network $G$\\
$T_I \equiv \tau_{I}$ & infectious period (assumed deterministic)   \\
 $T_L$ &latent period\\
$T_D$  & tracing delay  \\
\hline
\textbf{Parameter} & \textbf{Description} \\
\hline
$n$ & size of population (assumed large) \\

 $\mu = \mathbf{E}[D]$& mean degree\\
 $\sigma^2 = \mathbf{Var}[D]$& variance of degree distribution $D$\\
 $\beta_L$ & individual contact rate with each neighbour \\

 $\beta_G$ & rate of global contacts\\

$p_D$ &  probability that an individual who recovers is diagnosed \\

$p_M$ & probability that a diagnosed individual reports a given neighbour\\

$\pi_{A}$ & fraction of app-users \\
\hline
\end{tabular}
\label{tab:parameter}
\end{table}

\section{Early epidemic approximation in a large population}\label{sec:early_stage_approx}

\subsection{Approximation of the early epidemic without contact tracing}\label{sec:early_stage_no_ct}

It has been rigorously proven in \cite{ball2008network} that the process of infected individuals at the beginning of a SIR network epidemic with global contacts can be approximated by a two-type branching process with types $L$ and $G$  (type-$L/G$: infected by local/global contacts). This limiting result can be extended to hold also when a latent period is included, i.e., the SEIR epidemic model defined in Section \ref{sec:epi_model}. 

Suppose the number of initial susceptibles $(n-1)$ is large and the initial infective is infected from outside. During the early phase of an epidemic, the probability that an infective makes contact with a neighbour who has been infected (except its infector) is very small; it is also unlikely that an infective makes global contact with an already-infected individual. This suggests that all the local and global contacts are first made with distinct susceptibles. Consequently, the number of individuals locally (globally) infected by distinct infectives of the same type ($L$ or $G$) are independently and identically distributed. It follows that the process of infectives in the early stage of an epidemic can be approximated by a two-type branching process denoted by $E_{}(\beta_{L},\beta_{G},\tau_{I},T_{L},D)$ with type $L$ and $G$ described above.

\subsubsection{Basic reproduction number}
Let $M=(m_{ij})$ be the mean-offspring matrix of $E_{}(\beta_{L},\beta_{G},\tau_{I},T_{L},D)$, $m_{ij}$ denotes the mean number of type-$j$ individuals produced by a type-$i$ individual in the branching process, $i,j = L, G$. Any type of individual makes global contacts at rate $\beta_{G}$ during a constant period $\tau_{I}.$ It follows that
\begin{equation*}
    m_{LG} = m_{GG} = \beta_{G}\tau_{I}.
\end{equation*}
The degree of a given type-G infective is distributed according to $D$, and the probability that an individual infects a given neighbour is 
\begin{equation}\label{eq:prob_inf}
    p_{I}=1-e^{-\beta_{L}\tau_{I}}. 
\end{equation}
As a consequence, we get  
\begin{equation*}
    m_{GL} = \mu p_{I}.
\end{equation*}
On the other hand, a type-L infective has the so-called \textit{size-biased} degree distribution, denoted by $\Tilde{D}$ with $P(\Tilde{D}=k)=kp_{k}/\mu.$ More precisely, considering the infector of this type-L infective, an individual with degree $k$ is $k$ times more likely to be his/her neighbour than an individual with degree 1. Hence, a type-L infective has degree $k$ with probability proportional to $kp_{k}.$ At the beginning of the epidemic, a type-L infective has $\Tilde{D}-1$ susceptible neighbours (all except the infector). This implies that 
\begin{equation*}
    m_{LL} = \mathbf{E}[\Tilde{D}-1] p_{I}, 
\end{equation*}
where 
\begin{equation*}
    \mathbf{E}[\Tilde{D}-1] = \sum_{k=0}^{\infty}\frac{k^2p_k}{\mu} -1 = \frac{{\sigma}^2 + \mu^2}{\mu} -1 = \frac{{\sigma}^2 }{\mu} +\mu -1 . 
\end{equation*}
Let $R_{0}$ denote the largest eigenvalue of the mean-offspring matrix $M$ with elements $m_{ij}$ derived above, then 
\begin{equation}\label{eq:R0_general}
    R_{0}=\frac{1}{2}\Big(\beta_{G}\tau_{I}+\mathbf{E}[\Tilde{D}-1]p_{I}+\sqrt{(\beta_{G}\tau_{I}-\mathbf{E}[\Tilde{D}-1]p_{I})^2+4\beta_{G}\tau_{I}\mu p_{I}}\Big).
\end{equation}
According to multi-type branching process theory \cites{athreya_branching_1972,haccou2005branching}, a major outbreak can occur with positive probability if and only if $R_{0} >1$. We refer to $R_{0}$ as the \textit{basic reproduction number}.   

In addition, if $D$ follows a Poisson distribution, then $\mathbf{E}[\Tilde{D}-1]=\mu$, and the basic reproduction number is given by 
\begin{equation}\label{eq:R0}
    R_{0} = \beta_G\tau_{I} + \mu p_{I}
\end{equation}
with $p_{I}$ defined in Equation (\ref{eq:prob_inf}).

\subsection{Approximation of the early epidemic with manual and digital contact tracing}\label{sec:earlystage_comb}

During the early phase of the epidemic, assuming large $n$, it is very unlikely that an infectious individual will make contact with individuals who have already been infected by others (except for the infector on the network). The early epidemic with both types of contact tracing can be approximated by a limiting process, as described below. 

There are two types of individuals: app-users and non-app-users. While infectious, individuals have local infectious contacts with each susceptible neighbour at rate $\beta_{L}$ (and each neighbour is an app-user with probability $\pi_{A}$); have global infectious contacts with app-users at rate $\beta_{G}\pi_{A}$ and with non-app-users at rate $\beta_{G}(1-\pi_{A})$, respectively. An infected individual's infectious period starts after a random latent period $T_{L}$ and ends after an additional deterministic duration $\tau_{I}$. At the end of the infectious period, an individual is diagnosed with probability $p_{D}$, or else the individual recovers naturally without being diagnosed. A diagnosed non-app-user reports each of its network infectees independently with probability $p_M$, and each reported individual is traced after independent delays $T_D$. If an app-user is diagnosed, all of its app-using infectees are digitally traced immediately, and each of its local non-app-using infectees is reported with probability $p_M$ and manually traced after i.i.d. delays $T_D$. Traced individuals are tested and immediately isolate themselves if latent or infectious. Such traced individuals are however not contact traced further.

Below, we show that this limiting process can be characterized by a multi-type branching process. If we consider a sequence of epidemics indexed by population size $n$, we can then use a coupling argument (see \cite{ball_strong_1995} for details) to prove that the epidemic with manual and digital contact tracing converges almost surely to the limiting branching process (to be described below) on finite time intervals as $n\to\infty.$

We now characterize the limiting process as a multi-type branching process. To determine the different types, we start by considering four binary categories: app-user/non-app-user, infected by local/global contacts, infected with/without a digital CT link, and infected with/without a manual CT link. A \textit{manual CT link} between an infectee and his/her infector is only possible if the infector is diagnosed and the infectee is infected through the network. Conditioning on that, a manual CT link exists with probability $p_{M}.$ By having a \textit{digital CT link}, we mean that both infector and infectee are app-users and that the infector was diagnosed; so not having such digital CT link would correspond to that at least one of infector and infectee is a non-app-user or both of them are app-users, but the infector recovered naturally or was traced (implying that tracing will not happen). 

\begin{figure}[htb!]
    \centering
    \includegraphics[width=\textwidth]{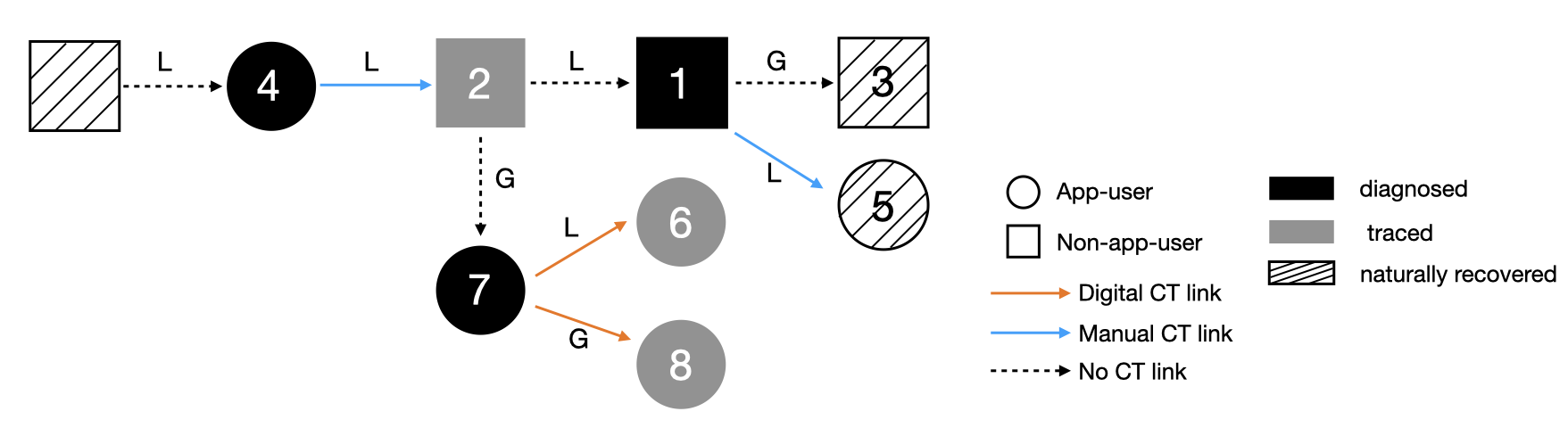}
    \caption{Example of an infection tree containing the eight types of individuals in Table \ref{tab:type_comb}. The circle nodes are app-users, and the squares are non-app-users. Nodes in black stand for ``diagnosed"; in white filled with diagonal lines, for ``naturally recovered"; and in grey, for ``traced". The arrows with ``L" are local (network) infections, while those with ``G" are global infections. }
    \label{fig:type_comb}
\end{figure}

In total, we would thus have $16$ types of individuals depending on the four binary categories. Among these $16$ types, $7$ are impossible: non-app-users having digital CT links (infected by local/global contacts with/without manual CT links), non-app-users that are infected by global contacts with no digital but manual CT links, app-users that are infected by global contacts with manual CT link and with/without digital CT links. This is because digital contact tracing can only occur between two app-users, and manual contact tracing happens only among local transmissions. Further, we can \textit{merge} two types: one is the app-user infected on the network with both digital CT link and manual CT link; the other is the same, but without the manual CT link. The reason is that if both types of contact tracing events occur, digital contact tracing always happens first, so manual contact tracing can be neglected. Hence, there are $16-7-1=8$ types of individuals (see an illustration in Figure \ref{fig:type_comb}).

Based on our assumptions, diagnosed individuals are infectious for a fixed period $\tau_{I}$, and contact tracing can only be triggered by diagnosed individuals. Following a similar argument in \cite{ball2015stochastic}, as a consequence of the deterministic infectious period $\tau_{I}$ and constant infection rates, the infection times of the infectees are independently uniformly distributed over the infectious period. Consider a diagnosed individual with $k$ infectees having CT links (either manual or digital). Label the $k$ infectees randomly and let $U_{i}$ be the time between infectee $i$ was infected until the infector is diagnosed, then $U_{1},..., U_{k}$ are hence independently and identically distributed as $U \sim U(0,\tau_{I})$. See Figure \ref{fig:ind_mct_link} for an illustration. If the infectious period $T_{I}$ is random,  $U_{i}$ will no longer be uniformly distributed and, more importantly, they will become dependent which disrupts the branching process approximation.

In summary, we can study the early stage of the SEIR network epidemic model with manual and digital contact tracing using an \textit{eight-type} branching process $E_{MD}$ with the types listed in Table \ref{tab:type_comb}. 

Let $M^{(MD)}=(m^{(MD)}_{ij})$ be the mean-offspring matrix (8-by-8) of $E_{MD}$, where the element ${m^{(MD)}_{ij}}$ represents the expected number of secondary infections of type $j$ produced by a single infected individual of type $i$, $i,j=1,\dots,8.$  The expressions of ${m^{(MD)}_{ij}}$ are given in Section \ref{sec:matrix_MD}. Further, let $R_{MD}$ be the largest eigenvalue of $M^{(MD)}$, it follows by the results from multi-type branching process theory \cites{athreya_branching_1972,haccou2005branching} that the branching process $E_{MD}$ dies out with probability 1 if $R_{MD} \leq 1$ and if $R_{MD} > 1,$ $E_{MD}$ grows beyond all limits with positive probability. As a consequence, the epidemic having $R_{MD} > 1$ may result in a major outbreak, while there will be a minor outbreak with probability 1 if $R_{MD} \leq 1.$ We thus refer $R_{MD}$ to as the \textit{effective reproduction number for the epidemic with manual and digital tracing}. 

\begin{table}[t!]
    \centering
        \caption{Description of the eight types in the limiting multi-type branching process $E_{MD}$}
    \begin{tabular}{|c|l|l|c|c|}
    \hline
    type & non-/app-user  & infected by & digital CT link & manual CT link\\
    \hline
    1 & \multirow{3}{*}{0 (non-app-user)} & 0 (local contacts) & \multirow{3}{*}{ 0 (No)} & 0 (No) \\
     \cline{3-3}\cline{5-5}
    2 &  &0 (local contacts)  &  & 1 (Yes) \\
     \cline{3-3}\cline{5-5}
    3 &  & 1 (global contacts) &  & 0 (No) \\
    \hhline{|=|=|=|=|=|}
    4 & \multirow{5}{*}{1 (app-user)} & \multirow{3}{*}{0 (local contacts)} & 0 (No)& 0 (No) \\
     \cline{4-5}
    5 &  &  &0 (No)  & 1 (Yes) \\
     \cline{4-5}
  6 &  &  & 1 (Yes) & 0/1 \\ 
   \cline{3-5}
    7 & &  \multirow{2}{*}{1 (global contacts)} & 0 (No) &\multirow{2}{*}{0 (No)} \\
    \cline{4-4}
    8 &  &  & 1 (Yes) & \\
    \hline
    \end{tabular}
    \label{tab:type_comb}
\end{table}

\subsubsection{Calculation of the mean-offspring matrix \texorpdfstring{$M^{(MD)}$}{}}\label{sec:matrix_MD}
We now derive all elements of the 8-by-8 mean offspring matrix, where the types are labelled as in Table \ref{tab:type_comb}. We start with offspring of types 1-3 being non-app-users. First, we note that digital CT links are exclusively formed between app-users. Consequently, non-app-users, categorized as types 1, 2, and 3, do not generate offspring with digital CT links, leading to the following equations:
\begin{equation}
    m^{(MD)}_{16}= m^{(MD)}_{18}= m^{(MD)}_{26}= m^{(MD)}_{28}= m^{(MD)}_{36}= m^{(MD)}_{38}=0.
\end{equation}

First, we note that each infectious contact an individual (app-user or not) makes, on the network or globally, is with an app-user with probability $\pi_{A}$ and with a non-app-user with probability $(1-\pi_{A})$. Secondly, we note that infected individuals who are not traced are diagnosed with probability $p_{D}$. If such a diagnosed individual is a non-app-user, its infectees through the network each have a manual CT link independently with probability $p_{M}$, and all global infectees lack CT links. If the diagnosed individual is an app-user, then all its app-using infectees have a digital CT link, its non-app-user network infectees have a manual CT link independently with probability $p_{M}$, and all its non-app-user global-infectees lack CT-links.

Consider now an individual without any CT link (types 1,3,4,7), which hence will not be traced. During the infectious period $\tau_{I},$ this individual generates, on average, $\beta_{G}\tau_{I}$ number of global infections. The number of local (network) infections depends on whether the individual was infected through the network or globally. If infected through the network, the individual has on average $\mathbf{E}[\Tilde{D}-1]p_{I}$  local infections, and on average $\mathbf{E}[D]p_{I}$ local infections if infected globally, where $p_{I}$ is defined in Equation (\ref{eq:prob_inf}).

These observations imply that an individual $i_0$ of type 1 (non-app-user infected locally without any CT link) has the following mean number of offspring. All global offspring lack CT links,  yielding $m^{(MD)}_{13}=\beta_{G}\tau_{I}(1-\pi_{A})$ and $m^{(MD)}_{17}=\beta_{G}\tau_{I}\pi_{A}$. If $i_{0}$ is diagnosed, each local offspring has a manual CT link with probability $p_{M}$, resulting in $m^{(MD)}_{12}=\mathbf{E}[\Tilde{D}-1]p_{I}p_{D}p_{M} (1-\pi_{A})$ and $m^{(MD)}_{15}=\mathbf{E}[\Tilde{D}-1]p_{I}p_{D}p_{M}\pi_{A}$. If not diagnosed, local offspring do not have CT links, leading to  $m^{(MD)}_{11}=\mathbf{E}[\Tilde{D}-1]p_{I}(1-p_{D}p_{M}) (1-\pi_{A})$ and $m^{(MD)}_{14}=\mathbf{E}[\Tilde{D}-1]p_{I}(1-p_{D}p_{M})\pi_{A}$. 

The expressions for non-app-users without CT links but infected globally (type 3) mirror those for type 1, but with $\mathbf{E}[\Tilde{D}-1]$ replaced by $\mathbf{E}[D]$ in the local infections. 

For app-users without CT links (types 4, 7) who will hence not be traced, global non-app-using infections lack CT links, resulting in $m^{(MD)}_{43}=m^{(MD)}_{73}=\beta_{G}\tau_{I}(1-\pi_{A}).$ Diagnosed app-users establish digital CT links with all app-using global offspring, leading to
$m^{(MD)}_{48}=m^{(MD)}_{78}=\beta_{G}\tau_{I}p_{D}\pi_{A}$ and 
\begin{equation}
    m^{(MD)}_{45}=m^{(MD)}_{75}=0.
\end{equation}
Undiagnosed app-users do not form CT links with their app-using global offspring, resulting in  $m^{(MD)}_{47}=m^{(MD)}_{77}=\beta_{G}\tau_{I}(1-p_{D})\pi_{A}$. For app-users infected by local contacts (type 4) and diagnosed, each local offspring is either linked through manual CT link with probability $p_{M}$ or through digital CT link if they are app-users, yielding $m^{(MD)}_{41}=m^{(MD)}_{11}$, $m^{(MD)}_{42}=m^{(MD)}_{12}$ and $m^{(MD)}_{46}=\mathbf{E}[\Tilde{D}-1]p_{I}p_{D}\pi_{A}$. Without diagnosis, local app-using offspring lack CT links, leading to $m^{(MD)}_{44}=\mathbf{E}[\Tilde{D}-1]p_{I}(1-p_{D})\pi_{A}.$ The structure of $m^{(MD)}_{7j}$ for app-users infected by global contacts (type 7) mirrors that of  $m^{(MD)}_{4j} (j=1,...,8)$, with $\mathbf{E}[\Tilde{D}-1]$ replaced by $\mathbf{E}[D]$.

\begin{figure}[htb!]
    \centering
    \includegraphics[width=.9\textwidth]{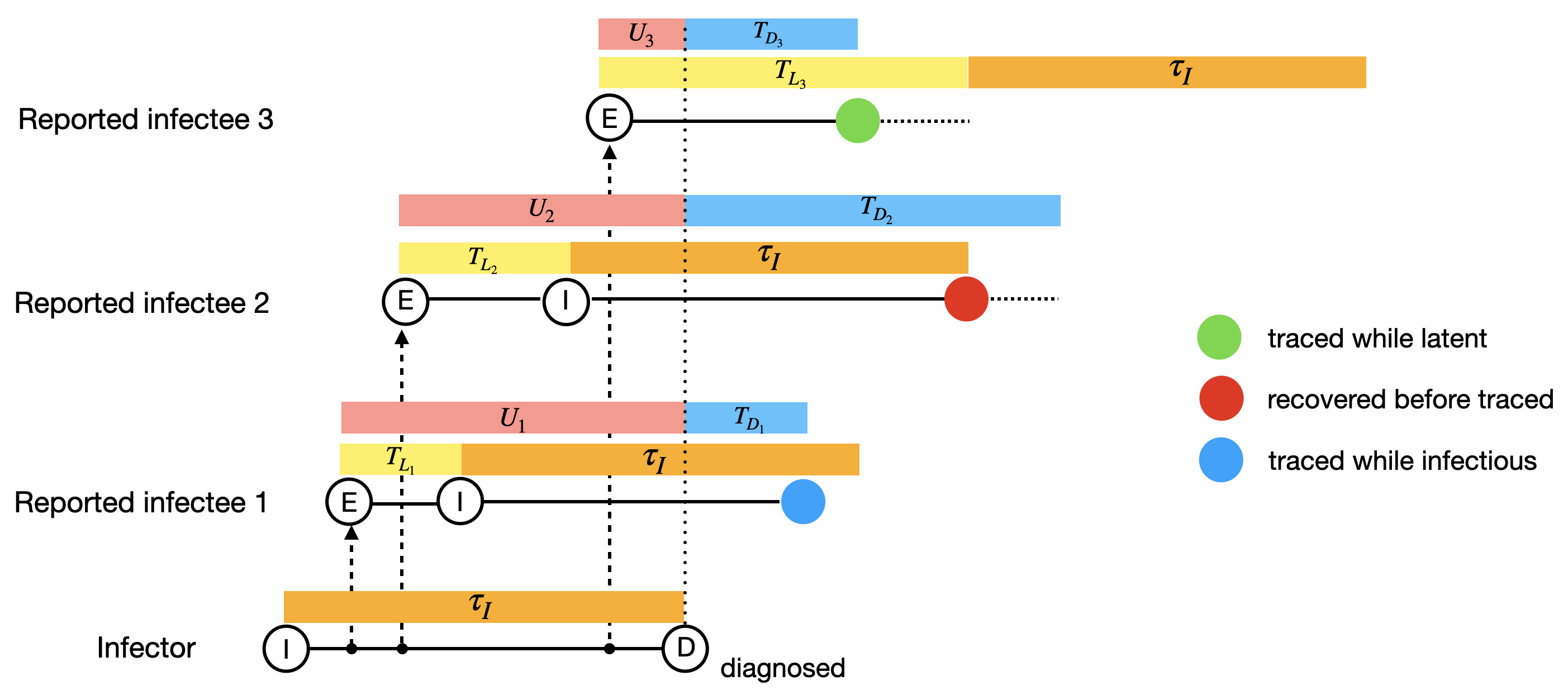}
    \caption{Example of the lifespans of three infectees (labelled in the order of time of infection) having manual CT link with same infector, where $U_{i} \overset{\mathrm{iid}}{\sim} U[0,\tau_{I}]$, $T_{L_{i}} \overset{\mathrm{iid}}{\sim} T_{L}$ and $T_{D_{i}}\overset{\mathrm{iid}}{\sim} T_{D}$, $i=1,2,3$. Infectee 1 is traced before the natural infectious period ends ($0<U+T_{D}-T_{L}<\tau_{I}$); infectee 2 is recovered before being traced ($U+T_{D}-T_{L}>\tau_{I}$); infectee 3 is traced during the latent period ($U+T_{D}-T_{L}<0$).}
    \label{fig:ind_mct_link}
\end{figure}

Next, we focus on individuals with manual CT links (type 2, 5). First, we let $i_{0}$ be of type 2, a non-app-user who has a manual CT link. The duration from $i_{0}$'s infection to the diagnosis of the infector follows a uniform distribution $U \sim U(0,\tau_{I})$. If $U+T_{D} \leq T_{L}$, $i_{0}$ is latent at the time of being traced and thus has no offspring. With probability 
\begin{equation}\label{eq:p_r_M}
    p^{(M)}_{R}:=\mathbf{P}(\tau_{I} \leq U-T_{L}+T_{D}),
\end{equation}
$i_{0}$ recovers naturally before the end of the tracing delay. In this case, $i_{0}$ generates offspring during $\tau_{I}$ and behaves like type-1 individuals (non-app-user without CT links). This implies that 
$m^{(MD)}_{22} = p^{(M)}_{R}m^{(MD)}_{12}$ and 
$m^{(MD)}_{25} = p^{(M)}_{R}m^{(MD)}_{15}$. On the other hand, if $0<U-T_{L}+T_{D}< \tau_{I}$, $i_{0}$ is infectious while being traced (see Figure \ref{fig:ind_mct_link} for an illustration showing the three scenarios). 

The probability of $i_{0}$ infecting a given neighbour and that $i_{0}$ is successfully contact traced is given by
\begin{equation}\label{eq:p_i_M}
    p^{(M)}_{I}:=\mathbf{E}[(1-e^{-\beta_{L}(U-T_{L}+T_{D})})1_{ \{0<U+T_{D}-T_{L}<\tau_{I}\}}].
\end{equation}
Consequently, such traced $i_{0}$ is expected to have $\mathbf{E}[\Tilde{D}-1]p^{(M)}_{I}$ number of local offspring, and if $i_{0}$ recovered before being traced, the average number of local offspring is $\mathbf{E}[\Tilde{D}-1]p_{I}p^{(M)}_{R}$. Only in the latter case could the local offspring have manual CT links. This leads to 
$m^{(MD)}_{21} = \mathbf{E}[\Tilde{D}-1]p^{(M)}_{I} (1-\pi_{A})+m^{(MD)}_{11}p^{(M)}_{R}$, 
$m^{(MD)}_{24} = \mathbf{E}[\Tilde{D}-1]p^{(M)}_{I}\pi_{A}+m^{(MD)}_{14}p^{(M)}_{R}$;
$m^{(MD)}_{22} = m^{(MD)}_{12}p^{(M)}_{R}$
and $m^{(MD)}_{25} = m^{(MD)}_{15}p^{(M)}_{R}$. 

For a traced $i_{0}$, the effective infectious period is $\max\{0,U-T_{L}+T_{D}\},$ during which the average number of global offspring is given by
$\beta_{G}\mathbf{E}[(U+T_{D}-T_{L})1_{ \{0<U+T_{D}-T_{L}<\tau_{I}\}}]$, where all the offspring lack CT links. The resulting expressions are 
$m^{(MD)}_{23} = \beta_{G}\mathbf{E}[(U+T_{D}-T_{L})1_{ \{0<U+T_{D}-T_{L}<\tau_{I}\}}](1-\pi_{A})+m^{(MD)}_{13}p^{(M)}_{R}$, and $m^{(MD)}_{27} = \beta_{G}\mathbf{E}[(U+T_{D}-T_{L})1_{ \{0<U+T_{D}-T_{L}<\tau_{I}\}}]\pi_{A}+m^{(MD)}_{17}p^{(M)}_{R}$, where 
\begin{equation}\label{eq:number_global_offspring_mct}
\begin{split}
    \beta_{G}\mathbf{E}[(U+T_{D} & -T_{L})  1_{\{0<U+T_{D}-T_{L}<\tau_{I}\}}]\\
    = &
    \frac{\beta_{G}}{2\tau_{I}}\mathbf{E}[(\tau_{I}+T_{D}-T_{L})^{2} 1_{\{-\tau_{I}< T_{D}-T_{L}< 0\}}]
     + \frac{\beta_{G}}{2\tau_{I}}\{\tau_{I}^{2} \mathbf{P}(0\leq T_{D}-T_{L}< \tau_{I})\\
     &-\mathbf{E}[(T_{D}-T_{L})^2 1_{\{0\leq T_{D}-T_{L}< \tau_{I}\}}]\}
\end{split}
\end{equation}
(the equality follows from calculation in \cite{ball2015stochastic} p.25).

We now assume $i_{0}$ is an app-user with a manual but no digital CT link (type 5), then all the app-using offspring would have digital CT links if $i_{0}$ is diagnosed before traced (with probability $p^{(M)}_{R}p_{D}$), This leads to $m^{(MD)}_{56}=p^{(M)}_{R}m^{(MD)}_{46}$, $m^{(MD)}_{58}=p^{(M)}_{R}m^{(MD)}_{48}$, and 
\begin{equation}
    m^{(MD)}_{55}=0.
\end{equation}
Despite this, $i_{0}$ generates the type-$j$ offspring for $j=1,2,3,4,7$ in the same way as for type 2 individuals, so $m^{(MD)}_{5j}=m^{(MD)}_{2j}$.  

\begin{figure}[htb!]
    \centering
    \includegraphics[width=.9\textwidth]{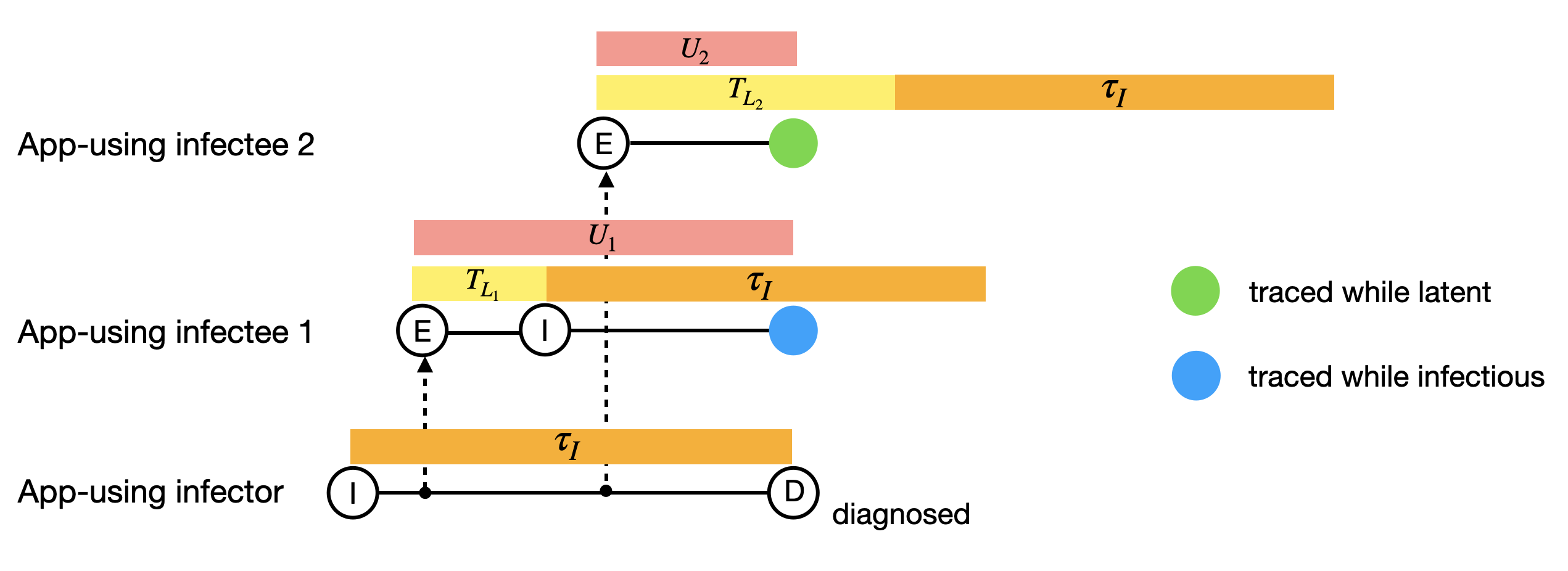}
    \caption{Example of the lifespans of two infectees (labelled in the order of time of infection) having digital CT link with same infector, where $U_{1}, U_{2}\overset{\mathrm{iid}}{\sim} U[0,\tau_{I}]$ and $T_{L_{1}}, T_{L_{2}} \overset{\mathrm{iid}}{\sim} T_{L}$. Infectee 1 is traced before the natural infectious period ends ($0<U-T_{L}<\tau_{I}$); infectee 2 is traced during the latent period ($U-T_{L}<0$).}
    \label{fig:ind_dct_link}
\end{figure}

Finally, we assume $i_0$ is an app-user with a digital CT link (type 6, 8). Due to the instantaneous nature of digital tracing, $i_{0}$ is immediately traced upon the diagnosis of the infector. It implies that the probability of $i_{0}$ recovering before being traced is given by
\begin{equation*}
   \mathbf{P}(\tau_{I} \leq U-T_{L}) = 0.
\end{equation*}
Consequently, $i_{0}$ could only generate offspring without CT links, implying that
\begin{equation*}
    {m^{(MD)}_{62}}={m^{(MD)}_{65}}={m^{(MD)}_{66}}={m^{(MD)}_{68}}=
{m^{(MD)}_{82}}={m^{(MD)}_{85}}={m^{(MD)}_{86}}={m^{(MD)}_{88}}=0.
\end{equation*}
Hence, with probability $1$, $i_{0}$ is traced and infectious of duration $\max\{0,U-T_{L}\}$ (see illustration in Figure \ref{fig:ind_dct_link}). The probability of $i_{0}$ infecting a given neighbor is thus given by
\begin{equation}\label{eq:p_i_D}
    p^{(D)}_{I}:=\mathbf{E}[(1-e^{-\beta_{L}(U-T_{L})})1_{\{U-T_{L}\geq 0\}}],
\end{equation}
which implies that $i_{0}$ has, on average, $\mathbf{E}[\Tilde{D}-1]p^{(D)}_{I}$ number of local offspring if infected through the network. This results in ${m^{(MD)}_{61}}= \mathbf{E}[\Tilde{D}-1]p^{(D)}_{I}(1-\pi_{A})$ and ${m^{(MD)}_{64}}= \mathbf{E}[\Tilde{D}-1]p^{(D)}_{I}\pi_{A}$. The elements $m^{(MD)}_{81}$ and $m^{(MD)}_{84}$ have the respective same structure as $m^{(MD)}_{61}$ and $m^{(MD)}_{64}$, but with $\mathbf{E}[\Tilde{D}-1]$ replaced by $\mathbf{E}[D]$. 

While infectious, $i_{0}$ also produces $\beta_{G}\mathbf{E}[(U-T_{L})1_{\{U-T_{L}\geq 0\}}]$ average number of global offspring, which yields ${m^{(MD)}_{63}}={m^{(MD)}_{83}}= \beta_{G} \mathbf{E} [(U-T_L) 1_{\{U-T_{L} \geq 0\}}] (1-\pi_{A})$ and ${m^{(MD)}_{67}}={m^{(MD)}_{87}}= \beta_{G} \mathbf{E} [(U-T_L) 1_{\{U-T_{L} \geq 0\}}] \pi_{A}$, where 
\begin{equation*}
    \beta_{G}\mathbf{E}[(U-T_{L})1_{\{U-T_{L}\geq 0\}}]=
    \frac{\beta_{G}}{2\tau_{I}}\mathbf{E}[(\tau_{I}-T_{L})^{2} 1_{\{T_{L}< \tau_{I}\}}], 
\end{equation*}
which is derived by setting $T_{D} \equiv 0$ in Equation (\ref{eq:number_global_offspring_mct}).

To summarize, the non-zero elements $m^{(MD)}_{ij}$ for $i,j=1,\dots,8,$ are as follows.
\begin{multicols}{2}
\noindent
    {\footnotesize\begin{align}
     {m^{(MD)}_{11}}&=\mathbf{E}[\Tilde{D}-1]p_{I}(1-p_{D}p_{M}) (1-\pi_{A})\label{eq: x_MD_11}\\
        {m^{(MD)}_{12}} &= \mathbf{E}[\Tilde{D}-1]p_{I}p_{D}p_{M} (1-\pi_{A})\label{eq: MD_12}\\
     m^{(MD)}_{13} &= \beta_{G}\tau_{I}(1-\pi_{A})\label{eq: MD_13}\\
        m^{(MD)}_{14} &= \mathbf{E}[\Tilde{D}-1]p_{I}(1-p_{D}p_{M}) \pi_{A}\label{eq: MD_14}\\
         m^{(MD)}_{15} &= \mathbf{E}[\Tilde{D}-1]p_{I}p_{D}p_{M}\pi_{A}\label{eq: MD_15}\\
         m^{(MD)}_{17} &= \beta_{G}\tau_{I} \pi_{A} \label{eq: MD_17}\\
         m^{(MD)}_{21} &= \mathbf{E}[\Tilde{D}-1](p_{I}p^{(M)}_{R}(1-p_{D}p_{M})
         &\notag\\&+p^{(M)}_{I}) (1-\pi_{A})\\
        m^{(MD)}_{22} &= \mathbf{E}[\Tilde{D}-1]p_{I}p^{(M)}_{R}p_{D}p_{M} (1-\pi_{A})\\
        m^{(MD)}_{23} &= \beta_{G}\{\mathbf{E}[(U+T_{D}-T_{L})1_{ \{0<U+T_{D}-T_{L}<\tau_{I}\}}]&\notag\\&+\tau_{I}p^{(M)}_{R}\}(1-\pi_{A})\\
m^{(MD)}_{24} &= \mathbf{E}[\Tilde{D}-1](p_{I}p^{(M)}_{R}(1-p_{D}p_{M})&\notag\\&+p^{(M)}_{I})\pi_{A}\\
m^{(MD)}_{25} &= \mathbf{E}[\Tilde{D}-1]p_{I}p^{(M)}_{R}p_{D}p_{M} \pi_{A}\\
         m^{(MD)}_{27} &= \beta_{G}\{\mathbf{E}[(U+T_{D}-T_{L})1_{ \{0<U+T_{D}-T_{L}<\tau_{I}\}}]
         &\notag\\&+ \tau_{I}p^{(M)}_{R}\}\pi_{A}\\
         {m^{(MD)}_{31}}&= \mathbf{E}[D]p_{I}(1-p_{D}p_{M}) (1-\pi_{A})\\
        {m^{(MD)}_{32}} &= \mathbf{E}[D]p_{I}p_{D}p_{M} (1-\pi_{A})\\
     m^{(MD)}_{33} &= \beta_{G}\tau_{I}(1-\pi_{A})\\
        m^{(MD)}_{34} &= \mathbf{E}[{D}]p_{I}(1-p_{D}p_{M}) \pi_{A}\\
         m^{(MD)}_{35} &= \mathbf{E}[{D}]p_{I}p_{D}p_{M} \pi_{A}\\
         m^{(MD)}_{37} &= \beta_{G}\tau_{I}\pi_{A}\\
         {m^{(MD)}_{41}}&= \mathbf{E}[\Tilde{D}-1]p_{I}(1-p_{D}p_{M}) (1-\pi_{A})\\     
{m^{(MD)}_{42}}&= \mathbf{E}[\Tilde{D}-1]p_{I}p_{D}p_{M} (1-\pi_{A})\\     
m^{(MD)}_{43} &= \beta_{G}\tau_{I}(1-\pi_{A})\\
m^{(MD)}_{44} &= \mathbf{E}[\Tilde{D}-1]p_{I}(1-p_{D})\pi_{A}\\
m^{(MD)}_{46} &= \mathbf{E}[\Tilde{D}-1]p_{I}p_{D} \pi_{A}\\
m^{(MD)}_{47} &= \beta_{G} \tau_{I} (1-p_{D}) \pi_{A}\\
m^{(MD)}_{48} &= \beta_{G} \tau_{I} p_{D} \pi_{A}\\
{m^{(MD)}_{51}}&= \mathbf{E}[\Tilde{D}-1](p_{I}p^{(M)}_{R}(1-p_{D}p_{M})&\notag\\&+p^{(M)}_{I})(1-\pi_{A})\\ 
 {m^{(MD)}_{52}}&= \mathbf{E}[\Tilde{D}-1]p_{I}p^{(M)}_{R}p_{D}p_{M} (1-\pi_{A})\\ 
  {m^{(MD)}_{53}}&= \beta_{G}\{\mathbf{E}[(U+T_{D}-T_{L})1_{ \{0<U+T_{D}-T_{L}<\tau_{I}\}}]&\notag\\&+ \tau_{I}p^{(M)}_{R}\}(1-\pi_{A})\\
 {m^{(MD)}_{54}}&= \mathbf{E}[\Tilde{D}-1](p_{I}p^{(M)}_{R}(1-p_{D}p_{M})&\notag\\&+p^{(M)}_{I})\pi_{A}\\
  {m^{(MD)}_{56}}&= \mathbf{E}[\Tilde{D}-1]p_{I}p^{(M)}_{R}p_{D}\pi_{A}\\
  {m^{(MD)}_{57}}&=\beta_{G}\{\mathbf{E}[(U+T_{D}-T_{L})1_{ \{0<U+T_{D}-T_{L}<\tau_{I}\}}]&\notag\\&+ \tau_{I}p^{(M)}_{R}\}\pi_{A}\\
  {m^{(MD)}_{58}}&=\beta_{G} \tau_{I}p^{(M)}_{R}p_{D}\pi_{A}\\
  {m^{(MD)}_{61}}&= \mathbf{E}[\Tilde{D}-1]p^{(D)}_{I}(1-\pi_{A})\\
{m^{(MD)}_{63}}&= \beta_{G} \mathbf{E} [(U-T_L) 1_{\{U-T_{L} \geq 0\}}] (1-\pi_{A})\\
  {m^{(MD)}_{64}}&=  \mathbf{E}[\Tilde{D}-1]p^{(D)}_{I}\pi_{A}\\
    {m^{(MD)}_{67}}&= \beta_{G} \mathbf{E} [(U-T_L) 1_{\{U-T_{L} \geq 0\}}] \pi_{A}\\
      {m^{(MD)}_{71}}&= \mathbf{E}[D]p_{I}(1-p_{D}p_{M}) (1-\pi_{A})\\
        {m^{(MD)}_{72}} &= \mathbf{E}[D]p_{I}p_{D}p_{M} (1-\pi_{A})\\
     m^{(MD)}_{73} &= \beta_{G}\tau_{I}(1-\pi_{A})\\
        m^{(MD)}_{74} &= \mathbf{E}[{D}]p_{I}(1-p_{D})\pi_{A} \\
         m^{(MD)}_{76} &= \mathbf{E}[{D}]p_{I}p_{D}\pi_{A}\\
         m^{(MD)}_{77} &= \beta_{G}\tau_{I}(1-p_{D})\pi_{A} \\
         m^{(MD)}_{78} &= \beta_{G}\tau_{I} p_{D}\pi_{A}\\
         {m^{(MD)}_{81}}&= \mathbf{E}[D]p^{(D)}_{I}(1-\pi_{A})\\
{m^{(MD)}_{83}}&= \beta_{G} \mathbf{E} [(U-T_L) 1_{\{U-T_{L} \geq 0\}}] (1-\pi_{A})\\
  {m^{(MD)}_{84}}&=  \mathbf{E}[D]p^{(D)}_{I}\pi_{A}\\
    {m^{(MD)}_{87}}&= \beta_{G} \mathbf{E} [(U-T_L) 1_{\{U-T_{L} \geq 0\}}] \pi_{A} \label{eq: x_MD_87}
\end{align}}
\end{multicols}

\noindent where $p_{I}$ is given by Equation (\ref{eq:prob_inf}), $p^{(M)}_{R}$ in Equation (\ref{eq:p_r_M}), $p^{(M)}_{I}$ in Equation (\ref{eq:p_i_M}) and $p^{(D)}_{I}$ in Equation (\ref{eq:p_i_D}).

\subsection{Approximation of the early epidemic with manual contact tracing only}
\label{sec:earlystage_mct}
In the previous section, we analyzed the initial stage of the epidemic with both manual and digital tracing and derived a reproduction number $R_{MD}$, determining whether there may be a major outbreak. In this section, our interest lies in the epidemic with manual tracing only. One straightforward approach would be to set $\pi_{A}=0$ in the elements of $M^{(MD)}$ and deriving its largest eigenvalue $R_{M}:=R_{MD}(\pi_{A}=0)$ which inherits the epidemic threshold property from $R_{MD}$. Hence, we refer $R_{M}$ as \textit{the effective reproduction number for the epidemic with manual tracing only}. 

Another way, still assuming a large $n$, we could approximate the early phase of the epidemic with manual tracing by a simpler \textit{three-type} branching process, denoted by $E_{M}$, with three types described in Table \ref{tab:type_mct}. Instead of considering four binary categories in Section \ref{sec:comb_model}, since digital tracing is not implemented (no app and digital CT links), we have only two binary categories: infected through the network or by global contacts, with manual CT link or not. Due to our assumption about manual tracing, those infected globally will never be reported (so no CT link). As previously discussed in Section \ref{sec:earlystage_comb}, the infection times of infectees with manual CT links are distributed uniformly and independently on $(0,\tau_{I})$. 
\begin{table}[htb!]
    \centering    
    \caption{The three types of limiting branching process $E_{M}$}
    \begin{tabular}{|c|l|c|}
    \hline
    type & infected by & manual CT link\\
    \hline
   1& 0 (local contacts) &  0 (No) \\
   2& 0 (local contacts)&   1 (Yes)  \\
   3& 1 (global contacts)&   0 (No)  \\
       \hline
    \end{tabular}

    \label{tab:type_mct}
\end{table}

Let $M^{(M)}=(m^{(M)}_{ij})$ be the mean-offspring matrix (3-by-3) of $E_{M},$ the expressions of ${m^{(M)}_{ij}}$ are presented as follows. The calculations are analogous to Section \ref{sec:matrix_MD}. 
\begin{multicols}{2}
{\footnotesize
\noindent
\begin{align}
m^{(M)}_{11} &= \mathbf{E}[\Tilde{D}-1]p_{I}(1-p_{D}p_{M}) \label{eq:x_M_11}\\
m^{(M)}_{12} &= \mathbf{E}[\Tilde{D}-1]p_{I}p_{D} p_{M} \\
  m^{(M)}_{13} &= \beta_{G}\tau_{I} \\
  m^{(M)}_{21} &= \mathbf{E}[\Tilde{D}-1] (p^{(M)}_{I}+ p_{I}p^{(M)}_{R}(1-p_{D}p_{M}))\\
   m^{(M)}_{22} &= \mathbf{E}[\Tilde{D}-1]p_{I}p^{(M)}_{R}p_{D}p_{M}\\
   m^{(M)}_{23} &=\beta_{G}\{\mathbf{E}[(U+T_{D}-T_{L})1_{\{0<U+T_{D}-T_{L}<\tau_{I}\}}]
   &\notag\\&+ \tau_{I}p^{(M)}_{R}\} \\
  m^{(M)}_{31} &= \mathbf{E}[D]p_{I}(1-p_{D}p_{M})\\
  m^{(M)}_{32} &= \mathbf{E}[D]p_{I}p_{D} p_{M}\\
  m^{(M)}_{33} &= \beta_{G}\tau_{I}, \label{eq:x_M_33}
\end{align}
}
\end{multicols}
\noindent with $p_{I}$ in Equation (\ref{eq:prob_inf}), $p^{(M)}_{R}$ in Equation (\ref{eq:p_r_M}) and $p^{(M)}_{I}$ in Equation (\ref{eq:p_i_M}).

We now show that the dominant eigenvalue of $M^{(M)}$ is identical to the one of $M^{(MD)}$ with $\pi_{A}=0$. It implies that the two approaches produce the same reproduction number $R_M$ for the epidemic with manual tracing. 

Setting $\pi_{A}=0$ in Equations (\ref{eq: x_MD_11})-(\ref{eq: x_MD_87}) gives that 
\begin{equation*}
{\footnotesize
M^{(MD)}(\pi_{A}=0) = 
\begin{pmatrix}
m^{(M)}_{11} & m^{(M)}_{12} & m^{(M)}_{13} & 0 & 0 & 0 & 0 & 0 \\
m^{(M)}_{21} & m^{(M)}_{22} & m^{(M)}_{23} & 0 & 0 & 0 & 0 & 0\\
m^{(M)}_{31} & m^{(M)}_{32} & m^{(M)}_{33} & 0 & 0 & 0 & 0 & 0\\
&&& 0 & 0 & 0 & 0 & 0\\
&&& 0 & 0 & 0 & 0 & 0\\
&A&& 0 & 0 & 0 & 0 & 0\\
&&& 0 & 0 & 0 & 0 & 0\\
&&& 0 & 0 & 0 & 0 & 0\\
\end{pmatrix}
}
\end{equation*}
with some 5-by-3 matrix $A$. 
Let $I_{k}$ be a $k$-by-$k$ identity matrix. Then the eigenvalues of $M^{(MD)}$ with $\pi_{A}=0$ solve the equation $det(M^{(MD)}(\pi_{A}=0) - \lambda I_{8})= 0$. Nevertheless, $M^{(MD)}(\pi_{A}=0) - \lambda I_{8}$ is a block-lower-triangle matrix (with 0's in the upper right corner). As a consequence, we have
$${\footnotesize det(M^{(MD)}(\pi_{A}=0) - \lambda I_{8})= det(M^{(M)} - \lambda I_{3})det(- \lambda I_{5})= (-\lambda)^{5} det(M^{(M)} - \lambda I_{3})}, $$
implying that the non-zero roots of $det(M^{(MD)}(\pi_{A}=0) - \lambda I_{8})=0$ coincide with those of $det(M^{(M)}- \lambda I_{3})=0$.
The largest eigenvalue of $M^{(MD)}$ with $\pi_{A}=0$  is hence identical with the largest eigenvalue of $M^{(M)} $.

\subsection{Approximation of the early epidemic with digital contact tracing only}
\label{sec:earlystage_dct}
Similarly, if our focus is epidemic with digital tracing only, we could analyze the threshold behaviour via setting $p_{M}=0$ in $M^{(MD)}$. The largest eigenvalue of $M^{(MD)}(p_{M}=0)$, denoted by $R_{D}$, is called the \textit{effective reproduction number for the epidemic with digital tracing only}. 

On the other hand, we can approximate the early phase of this special case by a \textit{six-type} branching process $E_{D}$ with types shown in Table \ref{tab:type_dct}, assuming a large population. Clearly, we do not have the binary category of having a manual CT link or not. The remaining binary categories are non-app-user/app-user, infected locally/globally, with digital CT link or not. Then we would have the $2^{3}=8$ types. However, since non-app-users will never be traced in this model, the two types where non-app-users have digital CT links are not possible. Hence we get $8-2=6$ types (see Table \ref{tab:type_dct}).

\begin{table}[htb!]
    \centering
        \caption{The six types in the limiting branching process $E_{D}$}
    \begin{tabular}{|c|l|l|c |c|}
    \hline
    type & non-/app-user   & infected by & digital CT link\\
    \hline
   1&   0 (non-app-user)  & 0 (local contacts)  & 0 (No) \\
   2&  0 (non-app-user)  & 1 (global contacts)& 0 (No)  \\
   3&  1 (app-user)   & 0 (local contacts)& 0 (No)  \\
   4& 1 (app-user) & 0  (local contacts)& 1 (Yes) \\
   5&  1 (app-user) & 1 (global contacts)& 0 (No) \\
   6&  1 (app-user) & 1 (global contacts)  & 1 (Yes)\\
   \hline
    \end{tabular}
    \label{tab:type_dct}
\end{table}

Let $M^{(D)}=(m^{(D)}_{ij})$ be the mean-offspring matrix (6-by-6) of $E_{D}$, the elements $m^{(D)}_{ij}$ are given as follows.

\begin{multicols}{2}
\noindent
    {\footnotesize\begin{align}
     m^{(D)}_{11} &= \mathbf{E}[\Tilde{D}-1]p_{I}(1-\pi_{A}) \\
     m^{(D)}_{12} &= \beta_{G}\tau_{I}(1-\pi_{A}) \\
     m^{(D)}_{13} & = \mathbf{E}[\Tilde{D}-1]p_{I}\pi_{A} \\
     m^{(D)}_{14}& = 0 \\
    m^{(D)}_{15}&=  \beta_{G}\tau_{I}\pi_{A}\\
    m^{(D)}_{16}&=  0 \\
    m^{(D)}_{21}&=  \mathbf{E}[D]p_{I}(1-\pi_{A})\\
    m^{(D)}_{22}&=  \beta_{G}\tau_{I}(1-\pi_{A})\\ 
    m^{(D)}_{23}&=  \mathbf{E}[D]p_{I}\pi_{A}\\ 
    m^{(D)}_{24}&= 0\\
    m^{(D)}_{25}&= \beta_{G}\tau_{I}\pi_{A}\\
    m^{(D)}_{26}&=  0 \\
    m^{(D)}_{31}&= \mathbf{E}[\Tilde{D}-1]p_{I}(1-\pi_{A}) \\
    m^{(D)}_{32}&= \beta_{G}\tau_{I}(1-\pi_{A})\\
    m^{(D)}_{33}&= \mathbf{E}[\Tilde{D}-1]p_{I}\pi_{A}(1-p_{D})\\
    m^{(D)}_{34}&= \mathbf{E}[\Tilde{D}-1]p_{I}\pi_{A}p_{D} \\
    m^{(D)}_{35}&= \beta_{G}\tau_{I}\pi_{A}(1-p_{D}) \\
    m^{(D)}_{36}&=  \beta_{G}\tau_{I}\pi_{A}p_{D}\\
    m^{(D)}_{41}&=\mathbf{E}[\Tilde{D}-1]p^{(D)}_{I}(1-\pi_{A})\\
    m^{(D)}_{42}&=\beta_{G}\mathbf{E}[(U-T_{L})1_{\{ U-T_{L} \geq 0\}}](1-\pi_{A})\\
    m^{(D)}_{43}&=\mathbf{E}[\Tilde{D}-1]p^{(D)}_{I}\pi_{A}\\
    m^{(D)}_{44}&=0\\
    m^{(D)}_{45}&=\beta_{G}\mathbf{E}[(U-T_{L})1_{\{ U-T_{L} \geq 0\}}]\pi_{A}\\
     m^{(D)}_{46}&=0\\
     m^{(D)}_{51} &=\mathbf{E}[D]p_{I}(1-\pi_{A})\\
     m^{(D)}_{52} &=\beta_{G}\tau_{I}(1-\pi_{A})\\
     m^{(D)}_{53}&=\mathbf{E}[D]p_{I}\pi_{A}(1-p_{D})\\
     m^{(D)}_{54}&=\mathbf{E}[D]p_{I}\pi_{A} p_{D}\\
     m^{(D)}_{55}&=\beta_{G}\tau_{I}\pi_{A}(1-p_{D})\\
     m^{(D)}_{56}&=\beta_{G}\tau_{I}\pi_{A} p_{D}\\
     m^{(D)}_{61}&=\mathbf{E}[D]p^{(D)}_{I}(1-\pi_{A})\\
     m^{(D)}_{62}&= \beta_{G}\mathbf{E}[(U-T_{L})1_{\{ U-T_{L} \geq 0\}}](1-\pi_{A})\\
     m^{(D)}_{63}&=\mathbf{E}[D]p^{(D)}_{I}\pi_{A}\\
     m^{(D)}_{64}&=0 \\
     m^{(D)}_{65}&=\beta_{G}\mathbf{E}[(U-T_{L})1_{\{ U-T_{L} \geq 0\}}]\pi_{A}\\
     m^{(D)}_{66}&=0
    \end{align}}
\end{multicols}

\noindent with $p_{I}$ in Equation (\ref{eq:prob_inf}) and $p^{(D)}_{I}$ in Equation (\ref{eq:p_i_D}).

As in the manual-only case, it holds that the two matrices, $M^{(D)}$ and $M^{(MD)}$ with $p_{M}=0$, have the same largest eigenvalue $R_{D}$ which we now show. 

Setting $p_{M}=0$ in Equations (\ref{eq: x_MD_11})-(\ref{eq: x_MD_87}) gives that 
\begin{equation*}
{\footnotesize
M^{(MD)}(p_{M}=0) = 
\begin{pmatrix}
m^{(D)}_{11} & 0 &m^{(D)}_{12}&m^{(D)}_{13}&0&m^{(D)}_{14}&m^{(D)}_{15}&m^{(D)}_{16}\\
*&0&*&*&0&0&*&0\\
m^{(D)}_{21} & 0 &m^{(D)}_{22}&m^{(D)}_{23}&0&m^{(D)}_{24}&m^{(D)}_{25}&m^{(D)}_{26}\\
m^{(D)}_{31} & 0 &m^{(D)}_{32}&m^{(D)}_{33}&0&m^{(D)}_{34}&m^{(D)}_{35}&m^{(D)}_{36}\\
*&0&*&*&0&0&*&0\\
m^{(D)}_{41} & 0 &m^{(D)}_{42}&m^{(D)}_{43}&0&m^{(D)}_{44}&m^{(D)}_{45}&m^{(D)}_{46}\\
m^{(D)}_{51} & 0 &m^{(D)}_{52}&m^{(D)}_{53}&0&m^{(D)}_{54}&m^{(D)}_{55}&m^{(D)}_{56}\\
m^{(D)}_{61} & 0 &m^{(D)}_{62}&m^{(D)}_{63}&0&m^{(D)}_{64}&m^{(D)}_{65}&m^{(D)}_{66}\\
\end{pmatrix},
}
\end{equation*}
where $*$ are some constants. Let $I_{k}$ be a $k$-by-$k$ identity matrix, the eigenvalues of $M^{(MD)}(p_{M}=0)$ solves $det(M^{(MD)}(p_{M}=0) - \lambda I_{8})=0$. 
By reordering the second and fifth rows to become the last two rows and similarly moving the second and fifth columns to the end, we transform the matrix $M^{(MD)}(p_{M}=0) - \lambda I_{8}$ into a block-lower-triangular form. Moreover, interchanging two rows leaves the value of the determinant unchanged, while swapping two columns only reverses the sign of the determinant. Consequently, we obtain 
$$ 0 = 
{\footnotesize
 det(M^{(MD)}(p_{M}=0) - \lambda I_{8})= det \begin{pmatrix}
M^{(D)} - \lambda I_{6} & \textbf{0}\\
B & - \lambda I_{2}\\
\end{pmatrix}
= \lambda^{2} det(M^{(D)} - \lambda I_{6}),}
$$
where $B$ is some 2-by-6 matrix, and the last equality is due to the determinant properties of a block-lower-triangular matrix and $det(- \lambda I_{2})=\lambda^{2}.$ As a result, the non-zero roots of $det(M^{(MD)}(p_{M}=0) - \lambda I_{8})=0$ align with those of $det(M^{(D)} - \lambda I_{6})=0,$ implying that $M^{(MD)}(p_{M}=0)$ and $M^{(D)}$ share the same largest eigenvalue.

In the next section, we illustrate our findings to better understand the effect of manual and digital contact tracing on the reproduction numbers. 

\section{Numerical Illustrations}\label{sec:numerical}
\subsection{Parameter setting}
In our numerical illustration, we focus on varying the parameters associated with contact tracing while keeping the other parameters fixed as follows. We assume the mean degree of the (social) network to equal $\mu = 5$, loosely motivated by the average household size in the EU being 2.3 (it is important to note, however, that we do not model households explicitly), and each person is assumed to meet on average 3 additional people regularly \cite{thurner2020network}. Further, we assume the deterministic infectious period to have duration $\tau_{I}= 5$ (days) \cite{kissler2020projecting}, and we also assume that the latency period is deterministic with duration $T_L \equiv 4$ (days) \cites{di2020impact,kissler2020projecting} having COVID-19 in mind. 
We assume the basic reproduction number equals $R_{0}=3$, but we also consider $R_{0}=1.5$, which could reflect a situation with physical distancing (contact rates are reduced accordingly).


In Section \ref{sec:numerical_diff_degree}, we see how the degree distribution $D$ influences the effectiveness of contact tracing by fixing the mean degree $\mu $ but varying its variance $\sigma^2$. In the other sections, we assume that $D \sim Poi(\mu)$, so $\mu=\sigma^2.$ Based on the expression of $R_{0}=\beta_{G}\tau_{I}+\mu p_{I}$ in Equation (\ref{eq:R0}), $\beta_G\tau_{I}$ can be viewed as the average number of global infections by a typical infective, and  $\mu p_{I}$ as the average number of local infections. We then define $\alpha = {\mu p_{I}}/{R_{0}}$ as the \textit{fraction of network transmission}. Due to that manual tracing is targeted only at infections on the network, $\alpha$ could have a non-negligible impact on the effectiveness of manual tracing. For this reason, at a given $R_{0}$, we choose contact rates $\beta_{L}$ and $\beta_{G}$ so that $\alpha=50\%$ (global and local infections have ``equal weights" \cites{ball1997epidemics, nande2021dynamics}) and $\alpha=75\%$ (more local than global infections) for comparison. Parameter values are summarized in Table \ref{tab:parameter_num}.

The code used to generate the figures presented in this section is publicly available at \cite{githublink}.

\begin{table}[htb!]
\centering
\caption{List of parameter values used in Section \ref{sec:numerical}}
\begin{tabular}{|l|l|}
\hline
\textbf{Parameter} & \textbf{Values}\\
\hline
Mean degree & $\mu = 5$ \cite{thurner2020network} \\
Latent period & $T_L \equiv 4$ days ($3.7$ days in \cite{di2020impact}, 4.6 days in\cite{kissler2020projecting})\\
Infectious period & $\tau_{I} = 5$ days \cite{kissler2020projecting}\\
Fraction of network transmission & $\alpha = 50\%$ or $75\%$ \\
Basic reproduction number & $R_{0}=3$ or $1.5$\\
\hline
\end{tabular}
\label{tab:parameter_num} 
\end{table}

\subsection{The effectiveness of manual and digital contact tracing}\label{sec:numerical_comb}
\begin{figure}[htb!]
\centering
 \includegraphics[width=.66\textwidth]{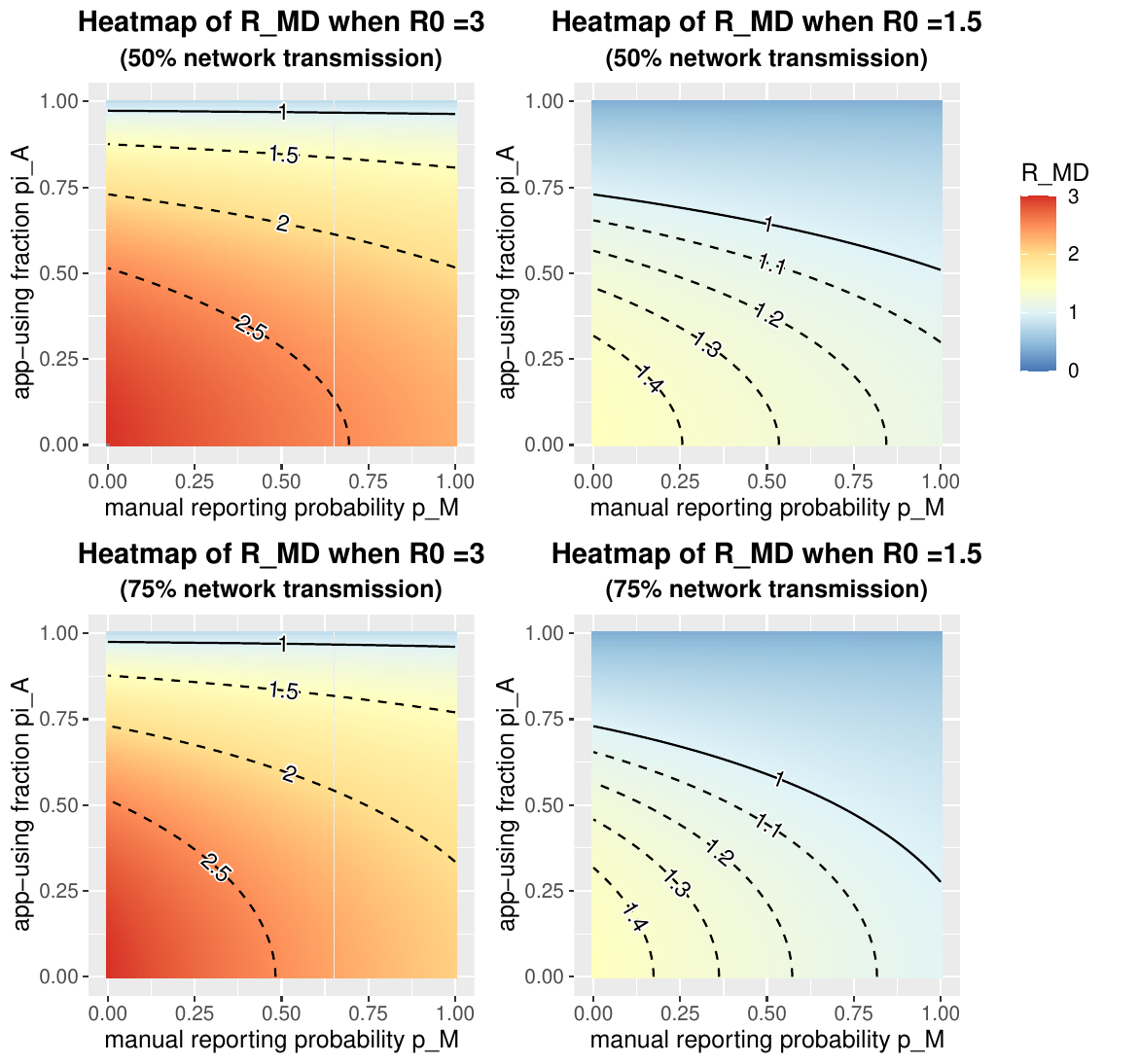}
\caption{Heatmaps of effective reproduction number $R_{MD}$ varying with $p_M \in [0,1]$ and $\pi_A \in [0,1]$ with fixed $p_{D} = 0.8$, $T_{D} \equiv 3,$ the contact rates $\beta_{L},\beta_{G}$ are chosen so that there are $50\%$ (top) and $75\%$ (bottom) network transmission given that $R_0=3$ (left panel) and $1.5$ (right panel) The black solid lines indicate where $R_{MD}=1$; the black dashed lines in the left panel are for $R_{MD}=1.5, 2, 2.5$ and in the right panel are for $R_{MD}=1.1,1.2,1.3,1.4$. 
}
\label{fig:heatmaps_MD}
\end{figure}

We start with investigating the effect of the manual reporting probability $p_M$ and the app-using fraction $\pi_A$ on $R_{MD}$ while keeping the probability of being diagnosed $p_{D} = 0.8$ (same as used in \cites{ball2015stochastic,kretzschmar2020impact}) and the tracing delay $T_{D} \equiv 3$ \cite{kretzschmar2020impact} fixed. As shown in Figure \ref{fig:heatmaps_MD}, $R_{MD}$ is monotonically decreasing in both $p_{M}$ and $\pi_{A}$ as expected, and the app-using fraction $\pi_A$ seems to be more influential in reducing $R_{MD}$ than $p_M$. However, when $R_{0}=3$, it is not possible to reduce $R_{MD}$ below 1 for realistic values on $p_{M}$ and $\pi_{A}$, meaning that contact tracing alone cannot prevent an outbreak. Even if $R_{0}=1.5$, a high fraction of app-users is needed to prevent an outbreak, or when most infections happen on the network ($\alpha=75\%$), effective manual contact tracing will also do the job with a moderate app-using fraction.

To illustrate the interplay between manual and digital tracing, we introduce $r_{M}, r_{D}$ and $ r_{MD}$ as the relative reductions in $R_{0}$ attributed to manual, digital and both types of contact tracing, respectively: 
\begin{equation*}
R_{M}=R_{0}(1-r_{M}), R_{D}=R_{0}(1-r_{D}), R_{MD}=R_{0}(1-r_{MD}).
\end{equation*}
If manual and digital contact tracing would have acted independently, then the reproduction number for the combined model would have been $R_{0}(1-r_{M})(1-r_{D}).$ This raises a question: How does $1-r_{DM}$ compare to the product $(1-r_{M})(1-r_{D})$? 

\begin{figure}[htb!]
\begin{subfigure}[b]{0.5\textwidth}
         \centering
 \includegraphics[width=\textwidth]{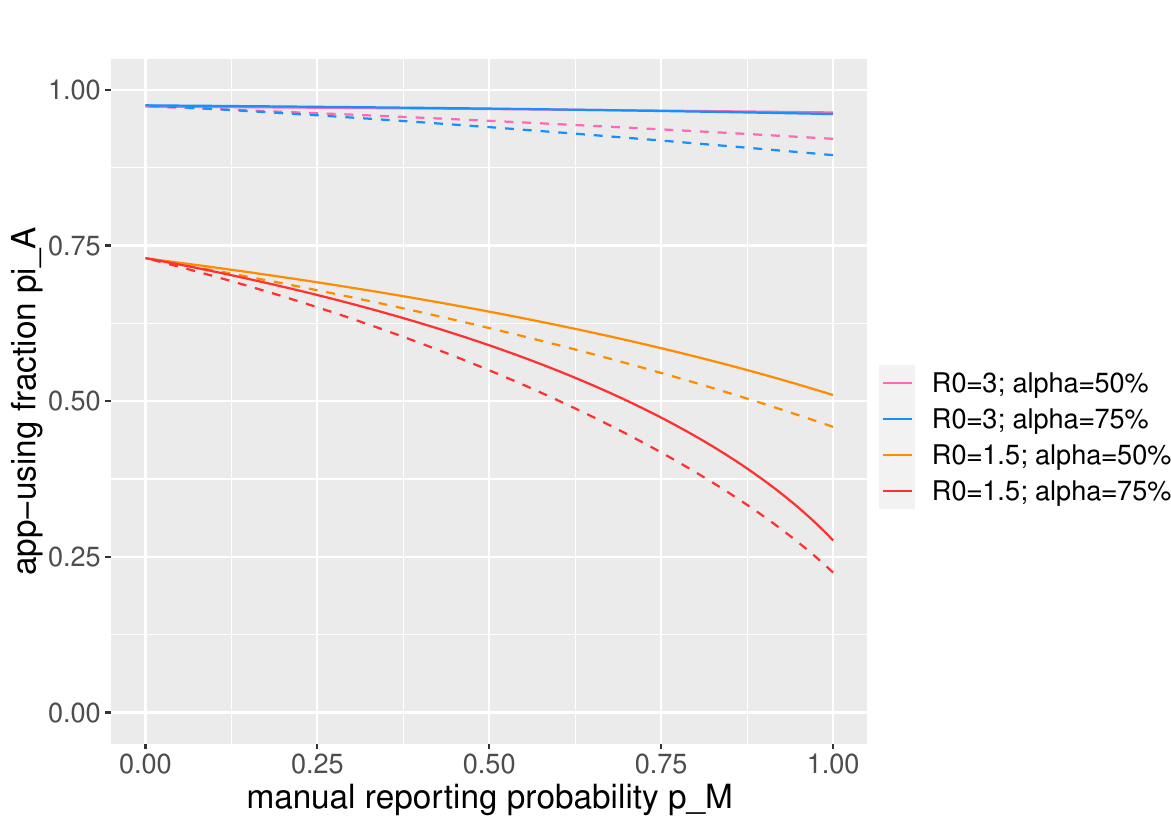}
 \caption{}
\label{fig:R_MD_critical}
     \end{subfigure}
\hfill
     \begin{subfigure}[b]{0.5\textwidth}
        \centering
    \includegraphics[width=\textwidth]{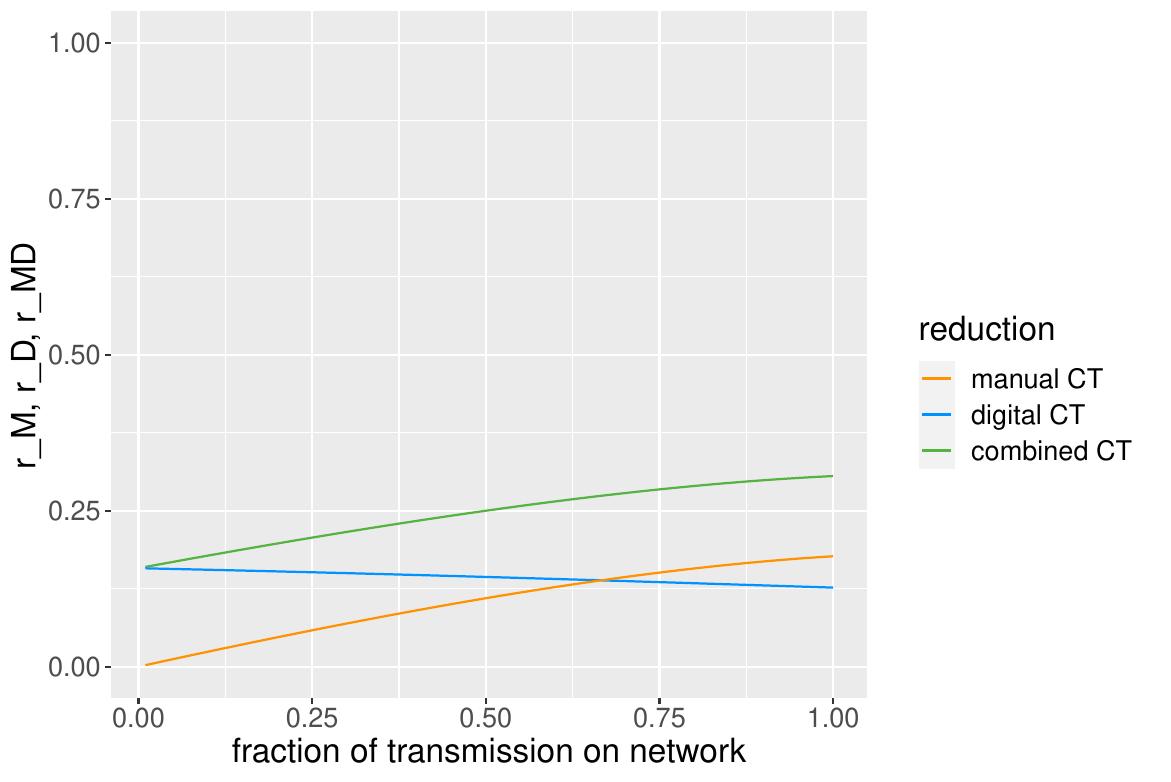}
    \caption{}
    \label{fig:r_all_alpha}
     \end{subfigure}
\caption{a): Plot of critical combination $(p_{M},\pi_{A})$ such that $R_{MD}=1$ (solid curves) and $R_{0}(1-r_{M})(1-r_{D})=1$ (dashed curves), where $p_{D} = 0.8$ and $T_{D} \equiv 3$ are fixed, and the contact rates $\beta_{L},\beta_{G}$ are chosen for $\alpha=50\%, 75\%$ and $R_0=3, 1.5$; b): Plot of the reductions $r_{M}, r_{D}$ and $ r_{MD}$ of $R_{0}$ against the fraction of network transmission $\alpha \in [0.01,1]$, where $R_{0}=3,$ $p_{D} = 0.8$, $T_{D} \equiv 3$ and $p_{M}=\pi_{A}=0.5$.}
\end{figure}

In Figure \ref{fig:R_MD_critical}, we plot the critical combinations of $(p_{M},\pi_{A})$ where $R_{DM}=R_0(1-r_{MD})=1$ against the scenarios where $R_0(1-r_{M})(1-r_{D})=1$. It shows that the true effect of combining manual and digital tracing is \textit{smaller} than the product of their separate effect! One reason could be that the \emph{timing} of an infection within the infectious period simultaneously influences the efficacy of \emph{both} manual and digital contact tracing: an infection at the beginning of the infectious period would make both digital and manual contact tracing have hardly any effect, and vice versa if the infection happens just before being diagnosed. In the latter case, one of the two types of contact tracing would be sufficient, and in the former, adding an additional type of contact tracing hardly helps. This argument suggests that the combined effect is smaller than had they acted independently, as observed in Figure \ref{fig:R_MD_critical}.

We then analyze the role of the network transmission fraction $\alpha$ on the efficiency of contact tracing (Figure \ref{fig:r_all_alpha}). Assuming that $p_{M}=\pi_{A}=0.5$, the effect of manual tracing $r_{M}$ increases with more network transmissions (as expected: manual contact tracing only happens among network transmissions), while the there is hardly any effect of digital tracing, in fact, $r_{D}$ actually decreases very slightly with $\alpha$. And the effectiveness of the combined tracing increases with $\alpha$ because manual contact tracing is more effective.

\subsection{The effectiveness of manual contact tracing}
\label{sec:numerical_mct}
\begin{figure}[htb!]
     \begin{subfigure}[htb!]{0.52\textwidth}
         \centering
 \includegraphics[width=\textwidth]{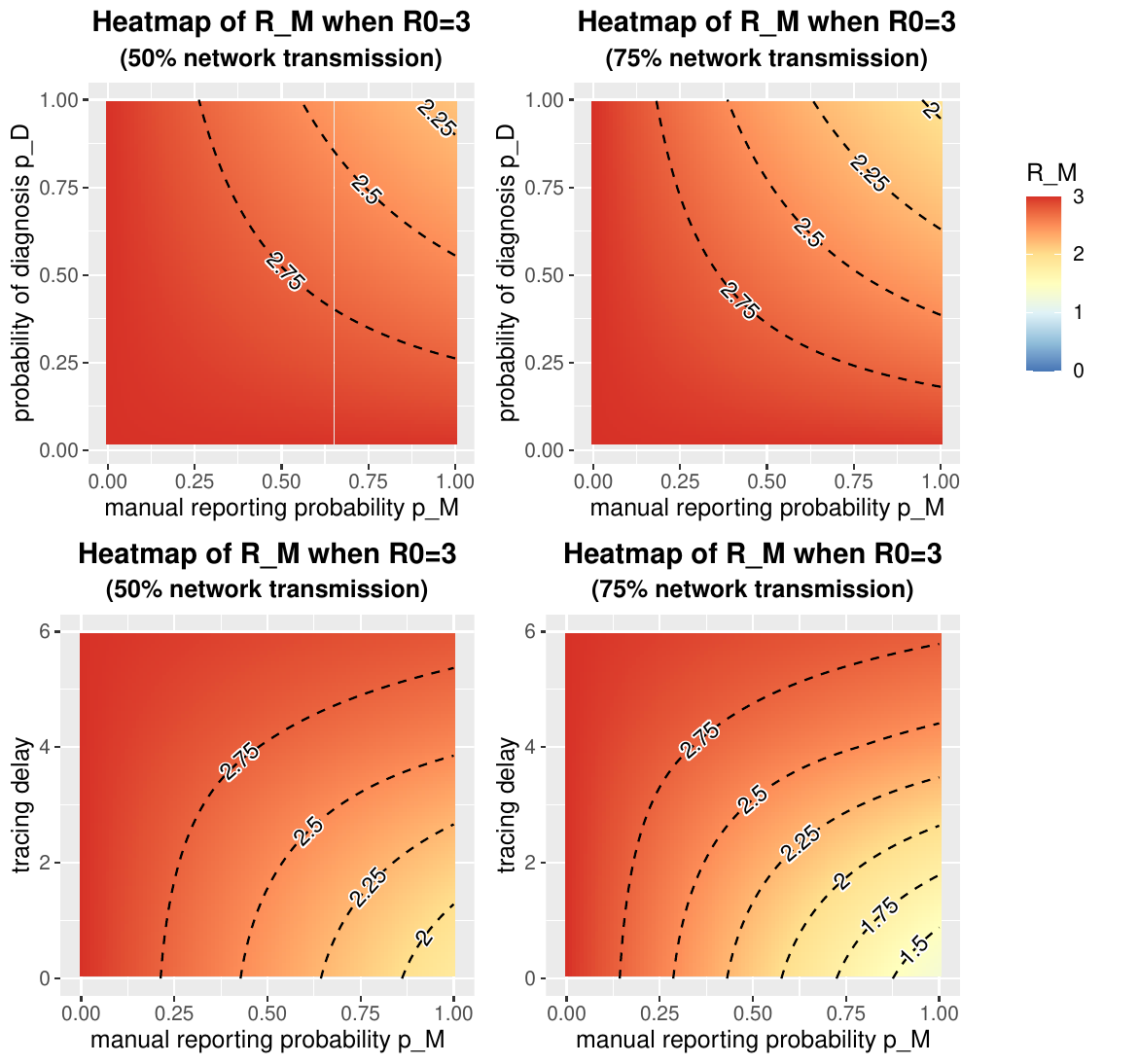}
 \caption{}
\label{fig:heatmap_M}
     \end{subfigure}
\hfill
     \begin{subfigure}[htb!]{0.52\textwidth}
         \centering
 \includegraphics[width=\textwidth]{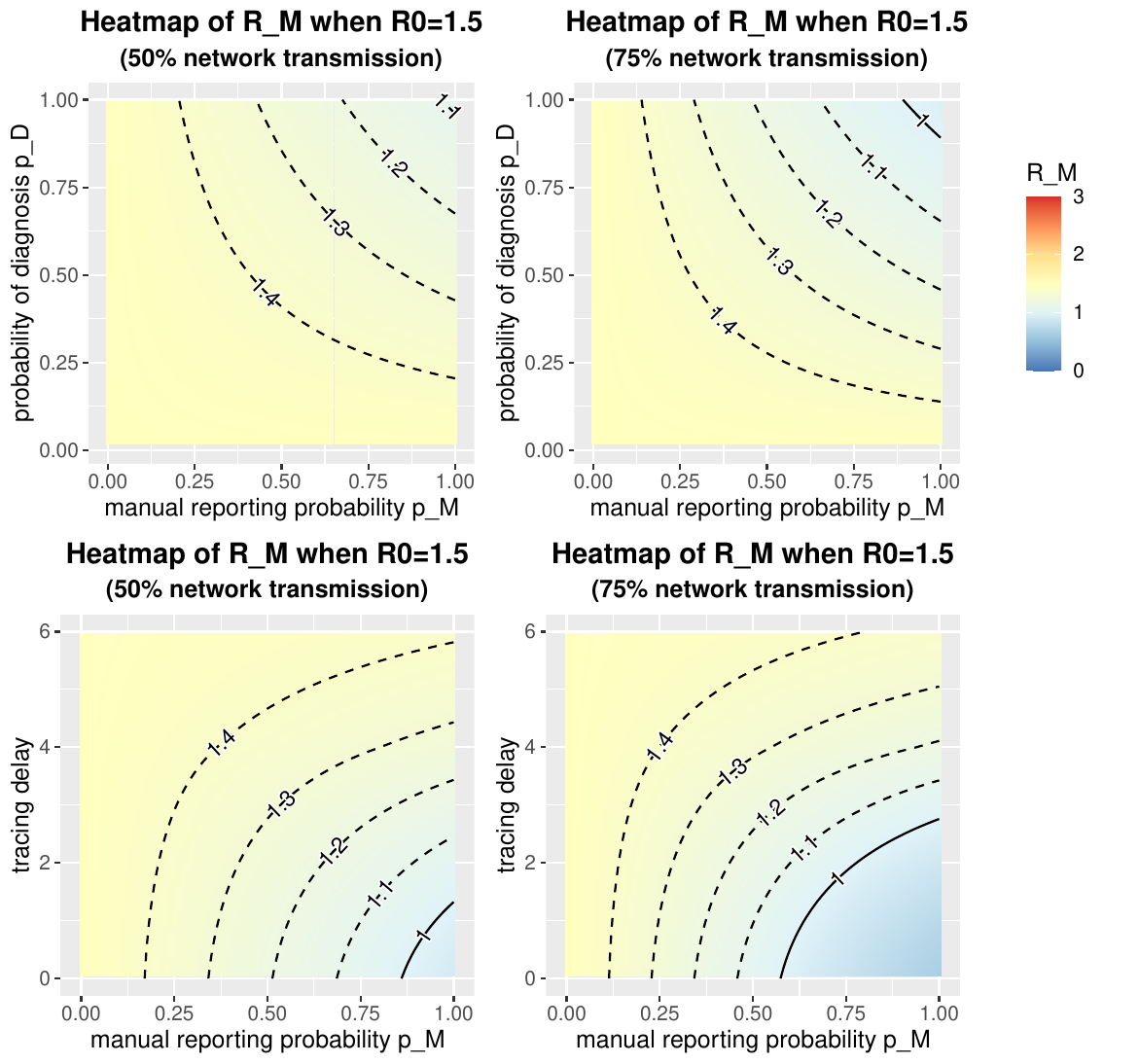}
 \caption{}
\label{fig:heatmaps_M_small_R0}
     \end{subfigure}
     
    \caption{Heatmaps of effective reproduction number $R_M$ varying with $p_M$ and $p_D$ in the upper panel when tracing delay $T_D \equiv 3$ days; the lower panel shows the heatmaps of $R_M$ varying with $p_M$ and deterministic tracing delay $T_D \in [0,6]$ while keeping $p_D=0.8$. Contact rates $\beta_{L},\beta_{G}$ are chosen so that there are $\alpha=50\%$ and $75\%$ transmissions on the network when a): $R_0=3$; b): $R_0=1.5$.}
    \label{fig:heatmaps_M}
\end{figure}

Next, we evaluate how the tracing delay $T_{D}$, the reporting probability $p_{M}$, and the diagnosis coverage $p_{D}$ influence the effectiveness of the manual contact tracing by means of the reproduction number $R_{M}$. Figure \ref{fig:heatmaps_M} shows that shortening the tracing delay, increasing the manual tracing probability $p_M$, and increasing the diagnosis probability $p_D$ all reduce the reproduction number $R_M$ as expected. Yet, we see from Figure \ref{fig:heatmap_M} that even perfect manual tracing (performed immediately and all contacts being reported) cannot reduce the reproduction number $R_{M}$ from an initial value of $R_{0}=3$ to below 1, primarily due to global transmissions not being traced. The extent of local transmission plays an important role in determining $R_{M}$, with more local transmission leading to lower $R_{M}$ under the same $(p_{D},p_{M})$ or $(T_{D},p_{M})$ configurations.

Figures \ref{fig:heatmaps_M_small_R0} reveals that for a baseline reproduction number of $R_{0}=1.5$, achieving $R_{M}=1$ is feasible with 
$p_{M}\approx 0.8$ and no tracing delay, provided that $\alpha=50\%$. If the network transmission increases to $75\%$, the tracing delay must be confined to a maximum of two days, or $p_{M}$ must be elevated to at least $0.6$ to reach $R_{M}=1$. 

To conclude, manual contact tracing is only effective if a large fraction of contacts are traced and the tracing delay is short, but it cannot alone reduce $R_{M}$ below 1 if $R_0$ is larger than e.g. 2.

\subsection{The effectiveness of digital contact tracing}
\label{sec:numerical_dct}
\begin{figure}[htb!]
\centering
 \includegraphics[width=.66\textwidth]{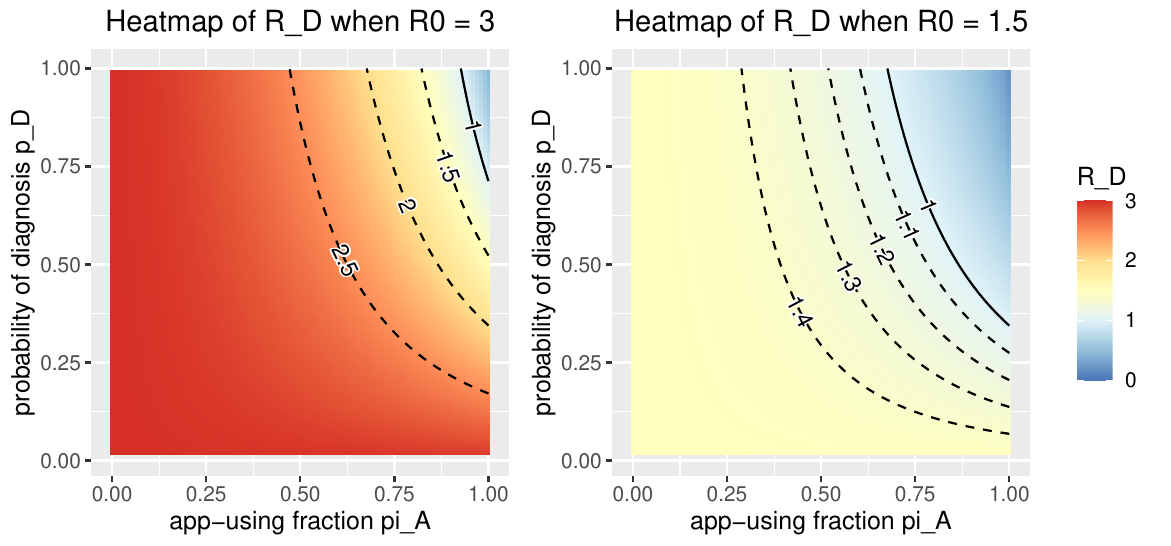}
\caption{Heatmaps of effective reproduction number $R_D$ varying with $p_D$ and $\pi_A$ when contact rates $\beta_{L},\beta_{G}$ are chosen so that there are $50\%$ transmission on network when $R_0=3$ and $1.5$. 
}
\label{fig:heatmaps_D}
\end{figure}

We then quantify the effect of the fraction of diagnosis $p_{D}$ and the fraction of app-users $\pi_{A}$ on the effectiveness of digital tracing (Figure \ref{fig:heatmaps_D}). It is observed that $R_{D}$ is monotonically decreasing with both  $p_D$ and  $\pi_A$ as expected. In particular, the app-using fraction $\pi_{A}$ affects $R_{D}$ more than $p_D$, the fraction of cases being diagnosed (by testing). With perfect testing ($p_{D}=1$), meaning that every infected individual is diagnosed, we need $\pi_A \geq 0.6$ for $R_{D} \leq 1$.

\begin{figure}[htb!]
\centering
\includegraphics[width=.66\textwidth]{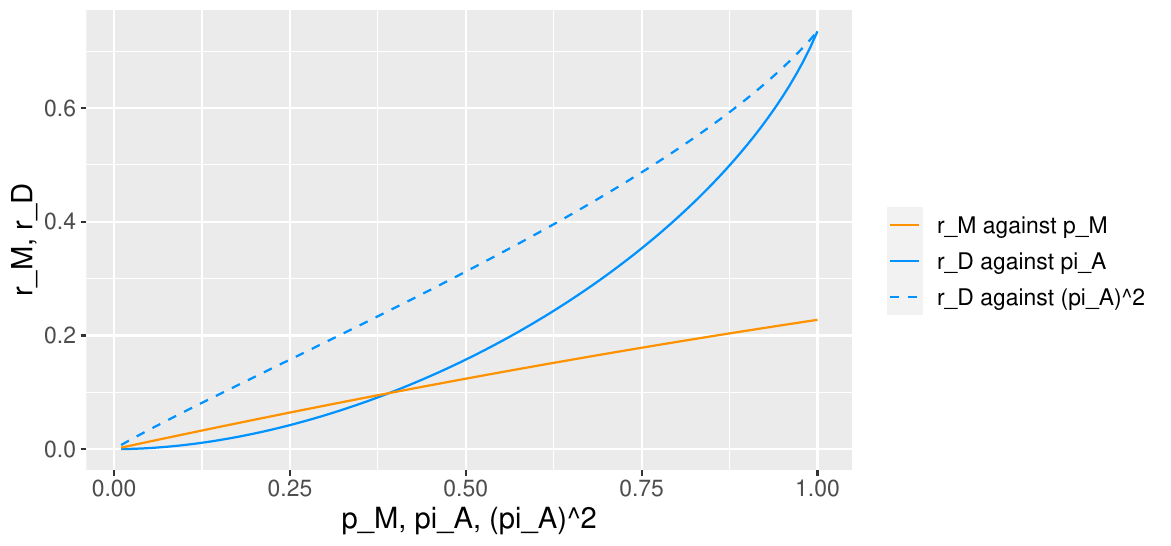}
\caption{Plot of the reduction $r_{M}$ and $r_{D}$ in the reproduction number with the effectiveness of contact tracing represented by $p_M$, $\pi_A$, and $\pi_{A}^2$, where $\beta_{L},\beta_{G}$ are chosen so that there are $50\%$ transmission on network when $R_0=3$, $p_D = 0.8$ and $T_D \equiv 3$.
}
\label{fig:plot_pi_square}
\end{figure}

We now numerically compare the effect of digital vs.\ manual contact tracing. For this comparison, we arbitrarily fix the delay in manual contact tracing to equal $T_D \equiv 3$ (deterministic) and the diagnosis probability to equal $p_D=0.8$. In Figure \ref{fig:plot_pi_square}, we compare $r_M$ and $r_D$ (the reduction in the reproduction numbers) when varying the app-using fraction $\pi_A$ and manual reporting probability $p_M$. In the figure, we plot the reduction of both reproduction numbers assuming $p_M=\pi_A$. It is seen that digital contact tracing outperforms manual contact tracing except when $p_M=\pi_A$ are small. The explanation is that manual tracing has a fairly long delay (3 days) for contact tracing to happen - the main advantage of digital tracing is that delays are dramatically shortened. Another disadvantage of manual tracing is that it only happens on contacts occurring on the social network (rarely can one name people on the bus or similar), whereas digital contact tracing happens on all close contacts between app-users. There is, of course, no reason to assume these $p_M$ and $\pi_A$ to equal each other, so a better comparison would be to compare the two curves at realistic values of $p_{M}$ and $\pi_A$, respectively.

A disadvantage with digital contact tracing is that it only happens for contacts where \emph{both} the infector and infectee are app-users, something which happens with probability $\pi_A^2$. This is why manual tracing is more effective for small values of $p_M=\pi_A$, and in most Western countries, the app-using fraction never became very big \cite{watch2020covid}. In the figure, we also plot the reduction $r_D$ as a function of $\pi_A^2$, which is seen to outperform manual tracing for all values of $\pi_A^2=p_M$.

\subsection{The effect of degree distribution}\label{sec:numerical_diff_degree}

\begin{figure}[htb!]
 \begin{subfigure}[htb!]{.5\textwidth}
        \centering
    \includegraphics[width=\textwidth]{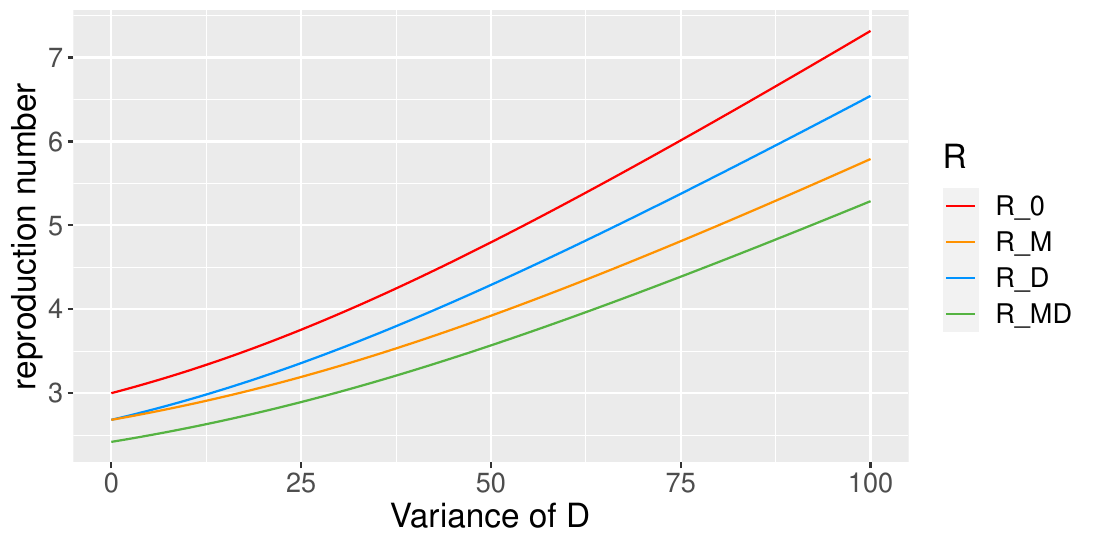}
    \caption{}
    \label{fig:R_CT_diff_variance}
     \end{subfigure}
        \hfill
\begin{subfigure}[htb!]{.5\textwidth}
        \centering
\includegraphics[width=\textwidth]{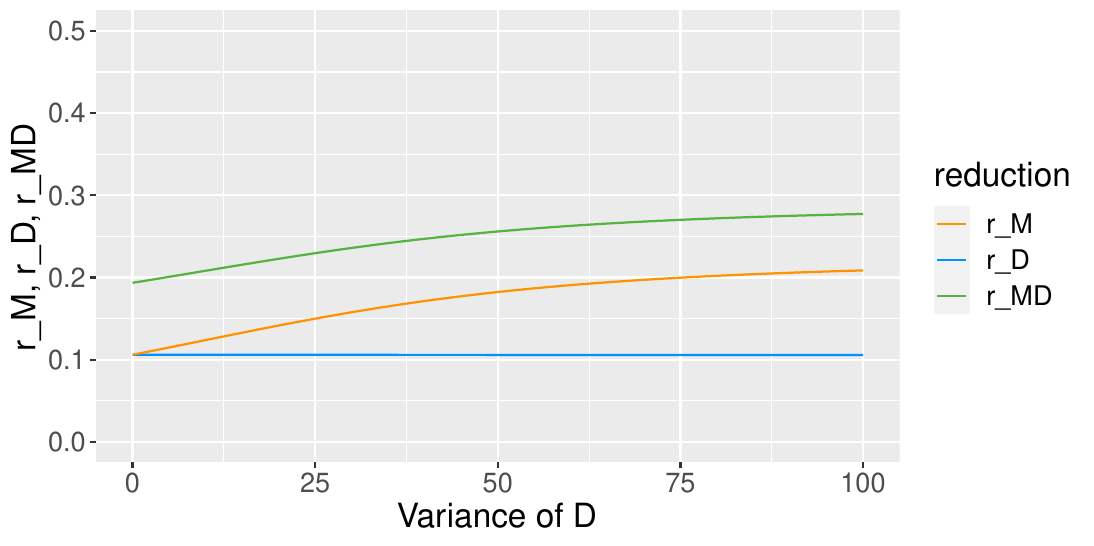}
 \caption{}
\label{fig:r_CT_diff_variance}
     \end{subfigure}
  
\caption{a): Plot of the reproduction numbers $R_{0}$, $R_{M}$, $R_{D}$ and $R_{MD}$ against $\sigma^2 \in [0,100]$. b): Plot of the corresponding reduction $r_{M}$, $r_{D}$ and $r_{MD}$ with $\sigma^2 \in [0,100]$. In both plots, we fix contact rates $\mu\beta_{L}=\beta_{G}$ so that $R_0(\sigma=0)=3$, as well as $p_{M}=0.5,p_{D}=0.8, T_{D}\equiv 3$ and $\pi_{A}\approx 0.41$ so that $R_{D}$ and $R_{M}$ ($r_{D}$ and $r_{M}$) start from the same value when $\sigma=0.$ 
}
\label{fig:diff_variance}
\end{figure}

We now examine the impact of the variance $\sigma^2$ of the degree distribution $D$ on the efficiency of contact tracing. As illustrated in Figure \ref{fig:diff_variance}, all four reproduction numbers $R_{0}$, $R_{M}$, $R_{D}$ and $R_{MD}$ are monotonically increasing with $\sigma^2$. Notably, the pairs $R_{0}$ and $R_{D}$ as well as $R_{M}$ and $R_{MD}$ seem to increase similarly with respect to $\sigma^2$, while the rate of increase for $R_{M}$ is comparatively slower than that for $R_{D}$. 

This could be explained as follows: Irrespective of  $\sigma^2$, the degree distribution of individuals identified via digital CT links is the same as that of the general infected individuals (both have the mixed degree distribution $\mathbf{P}(\text{local infection})\Tilde{D}+\mathbf{P}(\text{global infection})D$). However, individuals discovered through manual tracing tend to have a larger number of network contacts because they were infected through the network, so they have the size-biased degree distribution $\Tilde D$, which is affected more by an increase in $\sigma^2$ compared to $D$.

\subsection{Comparison with related results}
\label{sec:numerical_compare_proj2}
Finally, we compare the effectiveness of contact tracing as modelled herein with the contact tracing in our prior work \cite{zhang2022}. The model \cite{zhang2022} assumed both manual and digital tracing to be forward, backward, and iterative without delay; in the current work, we only consider forward contact tracing and non-iterative, and that manual contact tracing is equipped with a delay. The prior model hence represents a highly optimistic scenario, whereas the current model adopts a more conservative set of assumptions. Another difference is that the prior model also assumed a stochastic SIR epidemic in a homogeneously mixing population, whereas the current model allows for a latency period (SEIR model) and transmissions happening on a network as well as through homogeneous mixing.

\begin{figure}[htb!]
\centering
\includegraphics[width=\textwidth]{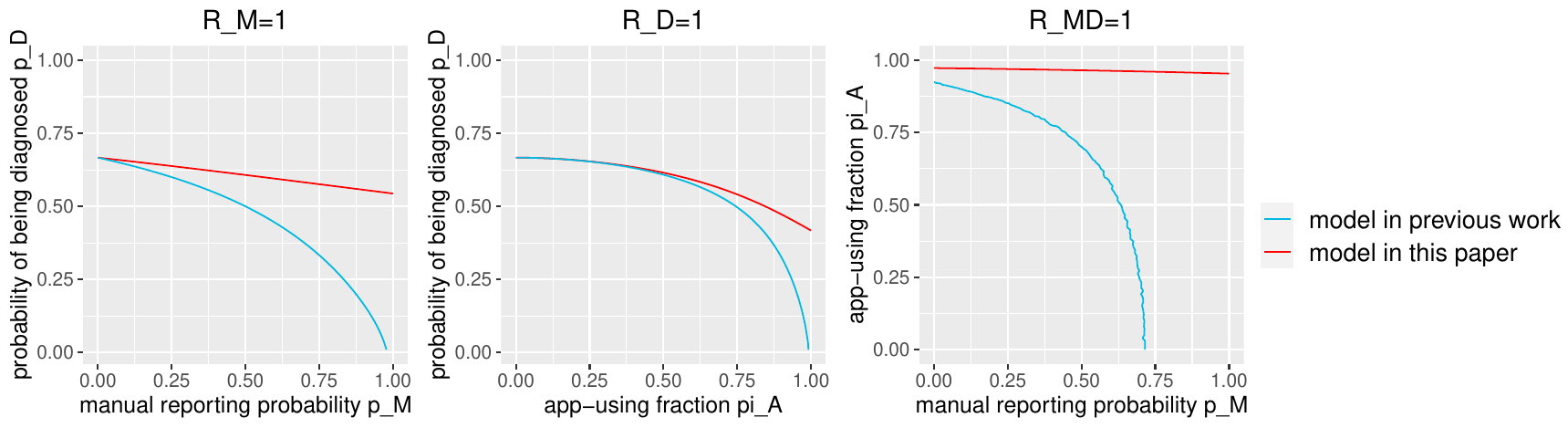}
\caption{Plots of two critical curves such that the reproduction number derived in \cite{zhang2022} and in this paper for the epidemic with manual-only (left), digital-only (middle) and combined model (right), equals 1, given that $T_D \equiv 3$ and in the combined model $p_{D}=0.8$.
}
\label{fig:compare}
\end{figure}

Figure \ref{fig:compare} illustrates the critical combination of parameters $(p_{D},p_{M})$, $(p_{D},\pi_{A})$ and $(p_{M},\pi_{A})$ such that the reproduction numbers $R_{M}, R_{D}$ and $R_{MD}$ equal 1 for both prior model in \cite{zhang2022} and current model. In the situations with manual/digital tracing only, for the prior model \cite{zhang2022}, we fix the rate of natural recovery $\gamma=1/5$ and the contact rate $\beta=3/5$ (so the basic reproduction number $\beta/\gamma$ in \cite{zhang2022} is 3). Further, we vary the fraction $\delta/(\delta+\gamma)$ ($\delta$ is the rate of diagnosis in \cite{zhang2022}), which aligns with our $p_{D}$ in the current model. Considering the average infectious period in the prior model \cite{zhang2022} is $1/(\delta+\gamma)$, we set $\tau_{I}$ in the present model to be $(1-p_{D}/\gamma)$. This ensures that both models share the same average infectious period. Moreover, we calibrate the contact rates $\beta_{L}$ and $\beta_G$ in the present model so that two critical curves for the manual-only and digital-only model (in the left and middle plots) start from the same point, which is done as follows. We first fix $\mu=5$, $\alpha=50\%$ and in the manual-only model $T_{D}\equiv3$. By obtaining the critical $p_{D}=2/3$ at which $R_{M}(p_{M}=0)$ in the previous model equals 1, we get a corresponding critical $\tau_{I}=5/3$. We then choose $\beta_{L}$ and $\beta_{G}$ such that $R_{M}(p_{M}=0)$ in the present model equals to 1. In the combined case, for the present model, we fix $p_{D}=0.8$ and contact rates $\beta_{L}$, $\beta_G$ are chosen such that $R_{0}=3$ with $\mu=5, \alpha=50\%$, while in the prior model \cite{zhang2022}, we set $\delta=p_{D}/\tau_{I}=0.8/5=4/25$ and $\gamma=1/5-\delta=1/25$ (so the mean infectious period $1/(\delta+\gamma)=\tau_{I}=5$), and $\beta=3/5$. 

It is not surprising that the overall effectiveness of the combined manual and digital tracing in the current study is lower, attributed to its conservative assumptions, including one-step, forward-only tracing, and delays in manual tracing. In particular, the efficacy of manual contact tracing is markedly smaller under these constraints. Conversely, for digital tracing, the critical curves of both models are similar until the app-using fraction exceeds $50\%$.

\section{Discussion}\label{sec:discussion}
This paper was concerned with an SEIR epidemic spreading on a (social) network as well as through random contacts, incorporating manual and digital contact tracing. The early stage of the epidemic, when combining manual and digital contact tracing, was shown to converge to an eight-type branching process as the population size $n\to\infty$. The reproduction number $R_{MD}$ was derived as the largest eigenvalue of the mean offspring matrix of the eight-type branching process. For scenarios limited to a single tracing strategy, the respective reproduction numbers $R_{D}$ and $R_{M}$ were obtained by setting the manual reporting probability $p_{M}$ or the app-using fraction $\pi_{A}$ to 0 in $R_{MD}$. Alternatively, the initial phase of the epidemic was approximated by a three-type branching process for manual tracing and a six-type branching process for digital tracing. 

Our numerical analysis shed light on the impact of various model parameters on the reproduction numbers. An increase in the app-using fraction  $\pi_{A}$ enhances digital tracing effectiveness, while manual tracing can be improved by either increasing reporting probability $p_{M}$ or reducing the tracing delays $T_D$. It was shown that $\pi_A$ plays a more significant role in controlling the epidemic than the manual reporting probability $p_M$. Another observation was that the relative efficacy of manual tracing vs digital contact tracing was amplified with increased network transmissions as well as when the variance of the network degree distribution is large since manual tracing only takes place for network contacts, and such contacts typically have a larger degree than globally infected individuals.

Despite these insights, several challenges remain to be solved for future work. A limitation of our model is that we restricted the infectious period to be deterministic so we could have a branching process approximation for the early epidemic, which seems harder in general. It would be interesting to extend the model to consider scenario where the infectious period is random (assuming the same mean). We hypothesize that this would \textit{increase} the efficacy of contact tracing. This is because individuals with longer infectious periods could potentially be traced well before the end of their infectious period, thereby significantly reducing the effect of such super-spreaders, but to prove this seems non-trivial. Further, we have only assumed one step instead of iterative tracing, while in most empirical cases, traced individuals that are infectious would be further questioned, thus possibly triggering manual and/or digital tracing, and so on. Moreover, we have not considered that app-using might have assortative nature, i.e., app-users' friends are more likely to use the app as well. It would be interesting to see how that influences the overall effectiveness of digital contact tracing. We also ignored the presence of clustering in the social networks, which would have consequences for the impact of both disease spreading and contact tracing \cites{kiss2005cluster,keeling_networks_2005}. Moreover, our current model does not include contact tracing within household settings. Household structures could potentially enhance the manual contact tracing process, since contacts within a household are more easily identified. An interesting problem not considered in the present paper is fatigue against contact tracing. An empirically observed problem with contact apps was that when incidence grew large, app-users would receive warnings very frequently, thus leading to fewer people following the instructions to self-isolate and test each time. It would be an interesting study to investigate how many tests individuals would be required to do for both types of contact tracing. We emphasize however that the current analysis focuses on the beginning of an epidemic outbreak, when incidence is low, implying that not many tests will be performed. In this paper, we focus mainly on contact tracing. There are other preventive measures that can effectively reduce the reproduction number, some of which can be considered within our model. For example, we can include social distancing by reducing the contact rates $\beta_{L}$ and $\beta_{G}$, leading to a smaller $R_{0}$. Additionally, implementing mass testing could increase the diagnosis probability $p_{D}$, thereby enhancing the effectiveness of contact tracing. 

Contrasting our model's ``pessimistic" assumptions (forward only and non-iterative contact tracing) with the over-optimistic scenarios of full, iterative and instantaneous contact tracing studied in \cite{zhang2022} indicates that real-world outcomes may lie somewhere in between. Moreover, our findings suggest that at the early stage of the epidemic, the actual combined effect of both manual and digital tracing is, unfortunately, smaller than the product of their separate effects. In conclusion, this paper contributes valuable perspectives on the interplay between manual and digital contact tracing in epidemic control and their respective contributions.

\subsection*{Acknowledgments}
T.B. and D.Z. thank the Swedish Research Council (grant 2020-04744) for financial support.

\subsection*{Declaration of generative AI and AI-assisted technologies in the writing process}
During the preparation of this work, D.Z. used ChatGPT in order to improve the readability. After using this tool, the authors reviewed and edited the content as needed and take full responsibility for the content of the publication.

\bibliography{reference}
\bibliographystyle{abbrvnat}

\end{document}